\documentclass[11pt]{article}
\usepackage{epsf,rotating,latexsym} 
\bigskip

\newcommand{\be}{\begin{equation}}
\newcommand{\ee}{\end{equation}}
\newcommand{\bd}{\begin{displaymath}}
\newcommand{\ed}{\end{displaymath}}
\newcommand{\bea}{\begin{eqnarray}}
\newcommand{\eea}{\end{eqnarray}}
\newcommand{\etal}{{\it et al.}}
\newcommand{\eg}{{\it e.g.}}
\newcommand{\ie}{{\it i.e.}}
\newcommand{\cf}{{\it cf.}}

\newcommand{\AU}{\,\hbox{AU}}
\newcommand{\mum}{\,\mu\hbox{m}}

\newcommand{\m}{\,\hbox{m}}
\newcommand{\km}{\,\hbox{km}}
\newcommand{\second}{\,\hbox{s}}

\newcommand{\g}{\,\hbox{g}}
\newcommand{\kg}{\,\hbox{kg}}
\newcommand{\V}{\,\hbox{V}}
\newcommand{\nT}{\,\hbox{nT}}

  \def\la{\mathrel{\mathchoice {\vcenter{\offinterlineskip\halign{\hfil
  $\displaystyle##$\hfil\cr<\cr\sim\cr}}}
  {\vcenter{\offinterlineskip\halign{\hfil$\textstyle##$\hfil\cr
  <\cr\sim\cr}}}
  {\vcenter{\offinterlineskip\halign{\hfil$\scriptstyle##$\hfil\cr
  <\cr\sim\cr}}}
  {\vcenter{\offinterlineskip\halign{\hfil$\scriptscriptstyle##$\hfil\cr
  <\cr\sim\cr}}}}}

\begin{document}

\begin{center}
{\Large \bf A Dust Cloud of Ganymede Maintained by Hypervelocity Impacts of                                          
Interplanetary Micrometeoroids} 

\bigskip

{\bf
        H.~Kr\"uger\footnote{{\em Correspondence to:} 
Harald Kr\"uger, Postfach 10\, 39\, 80, 69029 Heidelberg, Germany,
Tel.: +49-6221-516-563, 
Fax: +49-6221-516-324, 
E-Mail: Harald.Krueger@mpi-hd.mpg.de}, 
        A. V. Krivov$^{2,3}$,
        E.~Gr\"un$^1$
}

\begin{tabular}{ll}
1)& Max-Planck-Institut f\"ur Kernphysik, 69029 Heidelberg, Germany\\
2)& Max-Planck-Institut f\"ur Aeronomie, 37191 Katlenburg-Lindau, Germany\\
3)& On leave from Astronomical Institute, St.~Petersburg University,\\
  & 198904 St.~Petersburg, Russia
\end{tabular}

\end{center}

\normalsize

\begin{abstract}
A dust cloud of Ganymede has been detected by in-situ
measurements with the dust detector onboard the Galileo spacecraft.
The dust grains have been sensed at altitudes below five Ganymede radii
(Ganymede radius = $\rm 2,635\, km$).
Our analysis identifies the particles in the dust cloud surrounding Ganymede
by their impact direction, impact velocity, and mass distribution and
implies that they have been kicked up by hypervelocity
impacts of micrometeoroids onto the satellite's surface.
We calculate the radial density profile of the particles 
ejected from the satellite by interplanetary dust grains.
We assume the yields, mass and velocity distributions of the ejecta
obtained from laboratory impact experiments onto icy targets and
consider the dynamics of the ejected grains in ballistic
and escaping trajectories near Ganymede.
The spatial dust density profile calculated with interplanetary 
particles as impactors is consistent with the profile derived from the Galileo
measurements.
The contribution of interstellar grains as projectiles is negligible.
Dust measurements in the vicinities of satellites by
spacecraft detectors are suggested as a beneficial tool to obtain more
knowledge about the satellite surfaces, as well as 
dusty planetary rings maintained by satellites through the impact ejecta 
mechanism.
\end{abstract}

\medskip



\sloppy

\section{Introduction}

Since December 1995 the Galileo spacecraft has been on a bound orbit about 
Jupiter and performs regular close flybys of the Galilean satellites. 
With the dust detector onboard
({\it Gr\"un et al.,} 1992), at least three populations of dust have been identified 
in the Jovian system ({\it Gr\"un et al.,} 1997, 1998): 
1) Streams of ten-nanometre grains were detected throughout the Jovian 
magnetosphere and were recognizable even in interplanetary space out to
2~AU from Jupiter ({\it Gr\"un et al.,} 1993). Recently, it has 
been shown that these streams originate from Io, the 
ultimate source probably being the most powerful of its volcanic 
plumes ({\it Graps et al.,} 2000).
2) Bigger micrometre-sized particles form a tenuous dust ring 
between the Galilean satellites ({\it Colwell et al.,} 1998; 
{\it Thiessenhusen et al.,} 2000; {\it Krivov et al.,} 2000).
3) In the close vicinity of the Galilean satellites (i.e. within several 
satellite radii) strong sharp peaks were seen in the 
dust impact rate. The spatial location of the events, 
as well as the impact velocities, establish a genetic link between the 
detected dust grains and the satellites. Here we analyse this third
population of dust in the close vicinity of the moons. First results
have been published by {\it Kr\"uger et al.} (1999a). In this paper
we present the detailed data analysis and the
model for a dust cloud surrounding Ganymede.

The most important question addressed here is: what is the 
physical mechanism that ejected the grains off the moons?
Were these particles kicked up by hypervelocity impacts, presumably
of interplanetary micrometeoroids or tiny interstellar projectiles,
onto the surface of the satellites?
If this is true, we can treat the dust measurements made by Galileo
at the Galilean satellites as a unique natural impact experiment which,
for the first time, leads to the detection of the ejecta of hypervelocity
impacts in space. This process has been suggested as being responsible 
for maintaining both the Jovian ring ({\it Morfill et al.,} 1980; 
{\it Hor\'anyi and Cravens,} 1996; {\it Ockert-Bell et al.,} 1999, 
{\it Burns et al.,} 1999) 
and Saturn's E ring ({\it Hor\'anyi et al.,} 1992; {\it Hamilton and 
Burns}, 1994).
Our in-situ measurements in space give us a chance to gain more insight
into the fundamental properties of hypervelocity impacts, for
which the laboratory experiments on Earth still do not yield a
comprehensive picture.

Between December 1995 and April 2000 the Galileo
spacecraft had a total of 26 targeted encounters with all 4 
Galilean satellites: 4 with Io, 11 with Europa, 4 with Ganymede
and 7 with Callisto. In addition to the 4 Ganymede flybys 
which occurred in 1996 and 1997 
(encounters G1, G2, G7 and G8) two more 
are planned for the year 2000 (G28 and G29). 
Due to the spacecraft orientation, however, the 
population of grains existing in the close vicinity of Ganymede
will not be detectable during these two encounters.
Hence, the data set of dust particles detected with Galileo
during flybys at Ganymede will not be extended in the future.
Furthermore, the analysis of the data obtained from the whole Galileo
mission shows that the mass and velocity calibration
of the dust instrument is subject to a radiation-related 
ageing effect which affects data obtained after 1996 and which 
is still under investigation. 
For these two reasons, we here focus our attention on the dust grains
detected near Ganymede during the four close encounters with
this satellite.
Detailed investigations of the dust clouds surrounding 
Europa and Callisto will be the subject of a future paper.

We begin with a description of the Galileo data and their
processing (Section~2).
Section 3 presents the impact model that we use to interpret the data.
A comparison of the data and modelling results is made in Section 4.
Section 5 contains a summary and discussion of our findings.


\section{Dust impacts detected close to Ganymede}

\bigskip
{\em Dust impact detection}
\medskip

The dust detector onboard the Galileo spacecraft is a  
multi-coincidence impact ionisation dust detector  which
measures submicrometre- and micrometre-sized dust particles 
({\it Gr\"un et al.,} 1992; 1995). The instrument is 
identical with the dust detector onboard Ulysses.  

For each dust grain hitting the detector target three independent 
measurements of 
the impact-created plasma cloud are used to derive the impact speed 
$ v$ and  the mass $ m$ of the particle. 
The charge $ Q_I$ released upon impact on the target depends on 
the mass and the speed of the impacting grain according to:
$ Q_I \propto m \cdot v\,^{3.5}$.  The calibrated velocity range of the 
instrument is 2 to $\rm 70\,km\,s^{-1}$. The 
coincidence times of the three charge signals are used to classify each 
impact into one of four categories. 
Class~3 impacts have three 
charge signals, two are required for a class~2 and class~1 events, 
and only one for class~0.
Class 3, our highest class, are 
real dust impacts and class 0 
are noise events. Class~1 and class~2 events have been true 
dust impacts in interplanetary space  ({\it Baguhl et al.,} 1993, 
{\it Kr\"uger et al.,} 1999c).
 
Within about 15$\rm R_J$ distance from Jupiter
energetic particles from the Jovian plasma environment cause an 
enhanced noise rate in class~2 and the lower quality classes. 
By analysing the properties of the Io stream particles and 
comparing them with the noise events, the noise could be 
eliminated from the class~2 data ({\it Kr\"uger et al.,} 1999d), 
but class~1 shows signatures of being all noise in the Jovian 
environment. This algorithm, however, applies to the stream particles
and cannot per se be applied to other populations of grains.
Furthermore, the denoising technique 
uses statistical arguments and is applicable to large data sets. 
Individual dust impacts may be erroneously classified as noise and
vice versa. 
The analysis of the data sets from the flybys of all three satellites 
(Europa, Ganymede and Callisto) implies a slightly modified algorithm for the
rejection of noise events in the flyby data of Europa 
({\it Kr\"uger et al.,} 2000).

To check for a possible contamination by noise in the data set used for 
our analysis, we take all 
class~3 (dust impacts) and all class~2 events (dust and noise)
and compare this total number with the number of class~2 events
classified as noise by the standard algorithm ({\it Kr\"uger et al.,} 1999d)
and by the modified algorithm  applicable to the Europa data 
({\it Kr\"uger et al.,} 2000),
respectively. A total of 27\,\% of the events (10 out of 36)
would be classified as noise in the first case
and 14\,\% (5 out of 36) in the second one.
Even the higher possible noise
contamination in the first case does not affect the conclusions of
this paper. 
In addition, Ganymede's orbital radius is at
$\rm 15.0\,R_J$ from Jupiter at the edge of the noisy region 
where the noise contamination is relatively small anyway.

We therefore consider class~3 and all class~2 data for our analysis. 
It should be noted that there is no
physical difference between dust impacts categorised into 
class~2 and class~3. 

\begin{figure}
\epsfxsize=\hsize
\epsfbox{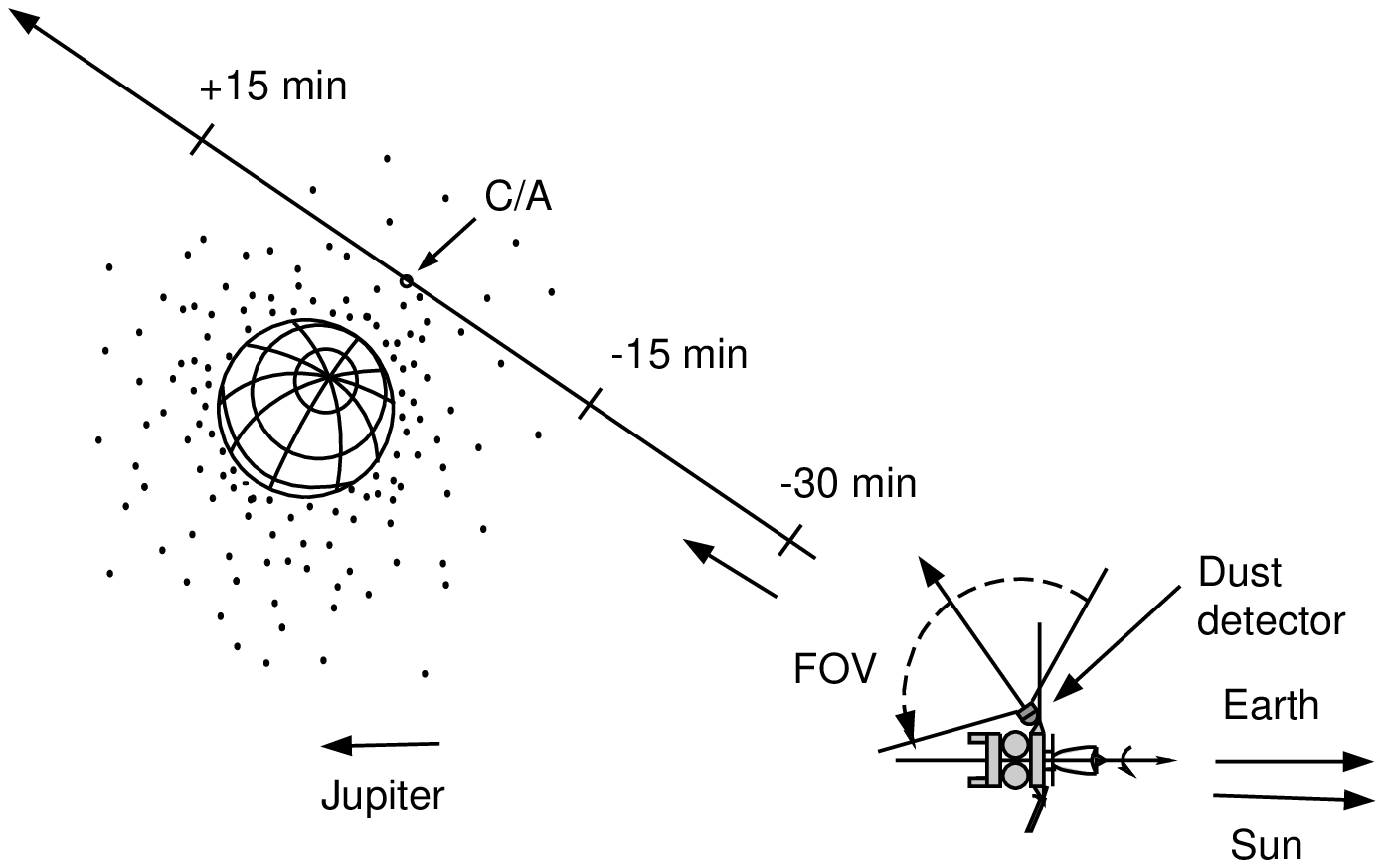}
\vspace{-3cm}
        \caption{\label{geometry}
Galileo's trajectory and geometry of dust detection during the G1 
Ganymede flyby. The Galileo spacecraft is sketched in an orientation it
was in during the flyby (see text for details). The directions to
Jupiter, Earth and Sun are shown. C/A indicates 
closest approach to Ganymede, FOV the field of view of the
dust instrument. The orientation of the dust instrument shown corresponds to 
a rotation angle $\Theta = 270^{\circ}$.  At $90^{\circ}$ rotation angle 
it points in the opposite direction.
}
\end{figure}

Galileo is a dual-spinning spacecraft, with an antenna that points 
antiparallel to the spacecraft positive spin axis. During most of the
orbital tour around Jupiter, the antenna points towards Earth. The
dust instrument is mounted on the spinning section of the spacecraft 
and its sensor axis is offset by an angle of $60^{\circ}$ from 
the spin axis (Fig.~\ref{geometry}; an angle of $55^{\circ}$ 
was erroneously stated earlier).

The rotation angle, $\Theta$, measures the viewing direction of the dust 
sensor at the time of particle impact. 
During one spin revolution of the spacecraft, the rotation angle scans
through $360^{\circ}$. Rotation angles for the Galileo dust instrument,
however, are reported opposite to that of the actual spacecraft rotation
direction. This is done to easily compare Galileo results with the 
dust detector data taken on the Ulysses spacecraft,  which, unlike Galileo, 
has the opposite spin direction. Zero degrees of 
rotation angle is taken when the dust sensor points close to the 
ecliptic north direction. At rotation angles of $90^{\circ}$ and 
$270^{\circ}$
the sensor axis lies nearly in the ecliptic plane (which corresponds
roughly to Jupiter's equatorial plane). The dust instrument has a 
$140^{\circ}$ wide field of view (FOV). Dust particles which arrive within 
$10^{\circ}$ of the positive spin axis can be sensed at all rotation 
angles, while those that arrive at angles from $10^{\circ}$ to 
$130^{\circ}$ from the positive spin axis can only be sensed over a 
limited range of rotation angles.

\bigskip
\centerline{\fbox{\bf Insert Figure \ref{geometry}}}
\medskip

Because Galileo's high-gain antenna did not open completely, 
the spacecraft has a very low data transmission capability. 
For the dust instrument this means that the full set of parameters measured 
during a dust particle impact (rotation angle, impact charges, 
charge rise times, etc.) can only be transmitted to Earth  for a
limited number of particles. During the times considered in 
this paper these limits are between one particle per minute 
and one particle per 21 minutes ({Kr\"uger et al.,} 2000). During periods of higher 
impact rates, the full set of parameters is not 
transmitted to Earth for all particles. All particles, however, 
are counted with one of 24 accumulators ({\it Gr\"un et al.,} 1995).
This allows for the determination of reliable impact rates 
during the satellite flybys.

\bigskip
{\em Impact direction}
\medskip

\begin{figure}
\epsfxsize=12.0cm
\epsfbox{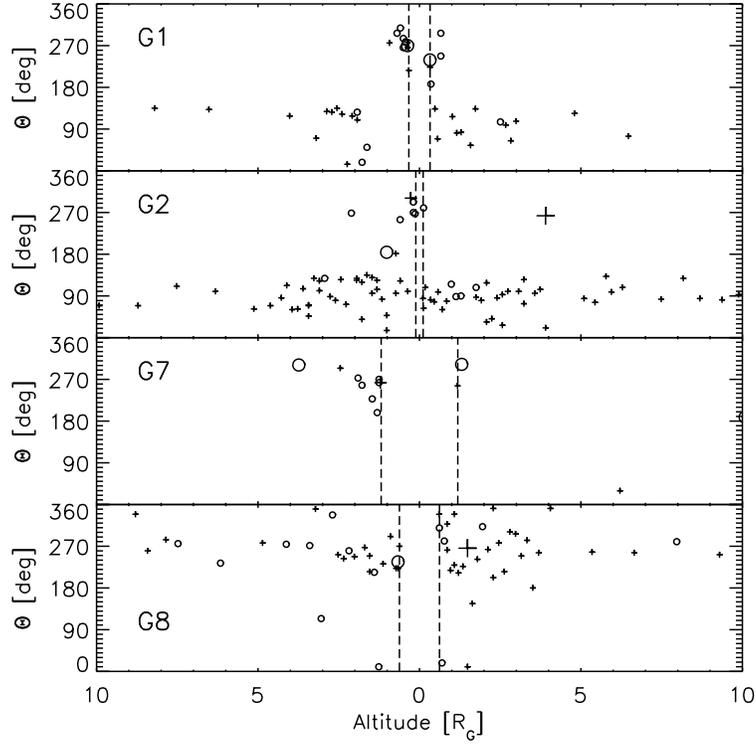}
        \caption{\label{rot_angle}
Sensor direction (rotation angle, $\Theta$) versus altitude of the Galileo 
spacecraft above the surface of Ganymede at the time of dust impact. 
Data are shown for all four Ganymede encounters (G1, G2, G7, G8).
The radius of Ganymede is $\rm R_G = 2,635\, km$.
The altitude range shown corresponds to a time interval of 2~h.
Each symbol indicates a dust particle impact and 
the size of the circle indicates the impact charge created by the 
particle ($ \rm 10^{-14}\,C \leq Q_{\,I} \leq 10^{-11}\,C$). 
Circles show particles with impact velocities below $\rm 10\,km\,s^{-1}$
and crosses show particles with higher speeds.
Galileo did not traverse the region between the vertical dashed lines.
}
\end{figure}

In Fig.~\ref{rot_angle} we show the impact direction (rotation 
angle) of those particles detected within about two hours around 
closest approach to Ganymede for which the complete 
set of measured impact parameters has been transmitted to Earth.
During the first three flybys (G1, G2, G7) particles
with rotation angles between $\rm 180^{\circ}$ and $\rm 360^{\circ}$ 
were strongly concentrated towards Ganymede. Most of them
were detected at altitudes below $\rm 2\,R_G$ (Ganymede radius 
$\rm R_G = 2,635\,km$). During the G1 and G2 flybys the large number 
of particles 
with rotation angles between $\rm 0^{\circ}$ and $\rm 180^{\circ}$
are streams of ten-nanometre-sized dust grains which have been 
detected throughout the Jovian system (see {\it Gr\"un et al.}, 1998,
for a detailed analysis). The direction from which
these dust streams were observed varied during Galileo's path 
through the Jovian system: when Galileo approached the
inner Jovian system rotation angles around $\rm 270^{\circ}$ were
observed. Shortly before closest approach to Jupiter the rotation
angle changed to $\rm 90^{\circ}$. Therefore, depending on when 
the satellite flyby occurred, stream particles approached from 
the corresponding direction: $\rm 90^{\circ}$ in the case of 
G1, G2 and G7, and $\rm 270^{\circ}$ during G8. The stream 
particles can, in general, be identified by their calibrated 
impact velocities as will be shown further below. 
The detection geometry is sketched in Fig.~\ref{geometry}.

At G1 and G2 an apparent concentration of particles 
with $\rm 0^{\circ} \leq \Theta \leq 180^{\circ}$ 
within about $\rm 5 \,R_G$ altitude is due to a higher data 
transmission rate for about an hour around Ganymede closest approach
(\cf\ Fig.~\ref{rot_angle}). During this time period the 
number of particles for which their complete information has been 
transmitted to Earth is increased by about a factor of ten.
At the G7 encounter almost all particles with 
$\rm 180^{\circ} \leq \Theta \leq 360^{\circ}$ were detected 
within $\rm 4\, R_G$ altitude. 
Only one stream particle was detected. 
During the G8 encounter the stream
particles approached with rotation angles between $\rm 180^{\circ}$ and 
$\rm 360^{\circ}$, and an apparent increase in the number of particles
within about $\rm 4 \,R_G$ altitude is again due to an enhanced 
data transmission rate.

\bigskip
\centerline{\fbox{\bf Insert Figure \ref{rot_angle}}}
\medskip

The first three panels of Fig.~\ref{rot_angle} (G1, G2 and G7) show
that particles detected with $\rm 180^{\circ} \leq \Theta \leq 360^{\circ}$
during the satellite flybys were concentrated towards 
Ganymede. They approached the dust sensor from a direction opposite 
to that of the dust stream particles. 

We have analysed the velocity 
vector of Ganymede in a coordinate system fixed to Galileo in order to 
find out if the grains detected with 
$\rm 180^{\circ} \leq \Theta \leq 360^{\circ}$ are compatible with a 
Ganymede origin. This shows that 
particles approaching the sensor from Ganymede's direction which have
a low velocity relative to Ganymede could be detected from a
direction corresponding to about $\rm 270^{\circ}$ rotation angle during all 
four encounters. Rotation angles of about $\rm 90^{\circ}$ are not 
compatible with the Ganymede direction. Taking into account the sensor 
field of view of $\rm 140^{\circ}$, particles detected during the 
G1, G2 and G7 encounters with rotation angles 
between $\rm 180^{\circ}$ and $\rm 360^{\circ}$ are compatible
with an origin from Ganymede itself. In the following we will 
call them Ganymede particles. 

A total number of  36 Ganymede particles has been identified below 
$\rm 10\,R_G$ altitude during these three encounters purely by their 
impact direction (\cf\ Tab.~1).
For our further analysis we use a cut-off altitude of 
$\rm 10\,R_G$ because this is close to the extension of the Hill 
sphere of Ganymede ($ R_H = \rm 12\,R_G$). For G1 and G2 the numbers of 
identified Ganymede particles are lower limits to 
the true numbers of detected grains 
because the complete set of parameters measured upon impact 
could be transmitted to Earth for only a fraction of all impacts. 
At G7, however, the complete set
of parameters for all particles detected within a two hour period 
around closest approach has been transmitted.

\bigskip
\centerline{\fbox{\bf Insert Table~1}}
\medskip

Ganymede particles from the G8 encounter cannot be identified by 
their impact direction alone because both stream particles and 
Ganymede particles approached the sensor from the same direction.
We therefore have to use an additional criterion 
to separate Ganymede particles at G8.

\bigskip
{\em Impact velocity}
\medskip

If the Ganymede particles truly originate from the satellite itself
and belong to a steady-state dust cloud they should impact the 
detector with roughly the velocity of  the spacecraft relative to 
the moon.
During all four flybys this encounter  velocity was close to
$\rm 8\,\km\second^{-1}$ (see Tab.~1) which is well above the 
detection threshold of the dust instrument for micrometre-sized 
grains ($\rm 2\,km\,s{-1}$). In contrast to the Ganymede particles,
stream particles have velocities in excess of $\rm 200\,\km\second^{-1}$ 
({\it Zook et al.,} 1996). Therefore, the impact velocity should be a good 
parameter to separate both populations of grains. 

\begin{figure}
\epsfxsize=12.5cm
\epsfbox{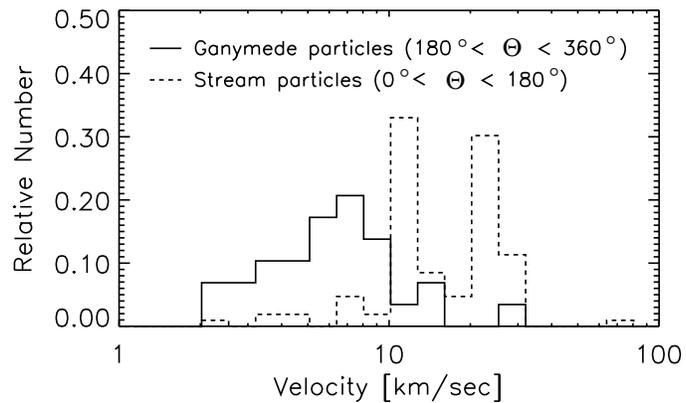}
        \caption{\label{velocities}
Impact velocities derived from the instrument calibration for 
dust particles detected at the
G1, G2 and G7 encounters below $\rm 10\,R_G$ altitude. The solid line shows the 
distribution for Ganymede particles ($\rm 180^{\circ} \leq \Theta \leq 
360^{\circ}$) and the dotted line that for stream particles
($\rm 0^{\circ} \leq \Theta \leq 180^{\circ}$). 
The mean velocity is $\rm 7.2 \pm 4.9\,\km\second^{-1}$. Only 
particles with a velocity error factor $\rm VEF < 6$ ({\it Gr\"un et al.,}
1995) have been considered (29 Ganymede particles).
}
\end{figure}

Calibrated impact velocities are derived from the rise-time of the impact 
charge signal by an empirically derived algorithm. Such impact 
velocities could be determined for 29 Ganymede particles 
detected below $\rm 10\,R_G$ altitude during the G1, G2 and G7 flybys
(velocity error factor $\rm VEF < 6$; {\it Gr\"un et al.,} 1995). 
In Fig.~\ref{velocities} we compare the calibrated impact velocities
of these Ganymede particles with those of the stream particles detected 
in the same time period. Because the distribution 
shows a bimodal structure, we can separate two statistical
subsets: nearly all of the stream particles 
have calibrated velocities in excess of $\rm 10\,\km\second^{-1}$, whereas 
Ganymede particles are slower. The true impact velocities of the stream
particles are much higher
than the upper limit of the calibrated velocity range of the dust instrument 
($\rm 70\,\km\second^{-1}$). Thus, the values derived from the instrument 
calibration are by far too low.

\bigskip
\centerline{\fbox{\bf Insert Figure \ref{velocities}}}
\medskip

The mean velocity of the 29 Ganymede particles in Fig.~\ref{velocities} is 
$\rm 7.2 \pm 4.9 \,\km\second^{-1}$ ($\rm 1\, \sigma$). Given a typical 
uncertainty for an individual velocity measurement of a 
factor of two, this value is in good agreement with the 
velocity of Galileo relative to Ganymede (about $\rm 8\,\km\second^{-1}$). 
It supports the picture that the particles 
detected belong to a dust cloud surrounding the moon. It also 
shows that the empirical velocity calibration of the instrument can be 
applied to the Ganymede particles, although one knows that it is not 
valid for the much smaller stream particles. Furthermore, and much more
important for our analysis here, the calibrated impact velocities can
be used to identify particles belonging to a dust cloud when such 
grains cannot be identified by their impact direction. It 
should be noted, however, that the two velocity distributions in 
Fig.~\ref{velocities} overlap which leads to some ambiguity in the 
identification of individual grains.

In Fig.~\ref{rot_angle} we have marked 
particles according to their calibrated velocities: those with
impact speeds below $\rm 10\,km\,s^{-1}$ are shown as circles, faster 
grains as crosses. For G8 (bottom panel) 
the majority of particles with $\rm 180^{\circ} \leq \Theta \leq 360^{\circ}$ have 
calibrated impact velocities above 
$\rm 10\,\km\second^{-1}$: they are
classified as stream particles and are rejected. Only 9 
particles with $\rm 180^{\circ}\leq \Theta \leq \rm 360^{\circ}$ remain 
below $\rm 5\,R_G$ altitude and a 
velocity below $\rm 10\,\km\second^{-1}$. Only particles with an 
altitude below $\rm 5\,R_G$ are considered for further analysis 
because we want to minimise the contamination by stream 
particles: a few stream 
particles in Fig.~\ref{velocities} have calibrated 
velocities below $\rm 10\,\km\second^{-1}$ and would erroneously be 
classified as Ganymede particles. Such a cutoff altitude is implied 
by the absence of Ganymede particles during the G7 flyby 
when the full data set for all particles has been transmitted.
The mean velocity of the 9
particles from G8 is $\rm 6.9 \pm 2.4\,\km\second^{-1}$ which is close 
to the value derived for the three earlier encounters.

At the G8 encounter the identified Ganymede particles show a slight 
concentration towards the satellite. Together 
with the 36 particles identified by their impact direction 
during the three earlier flybys, we have identified a total set of 45 Ganymede 
particles from all four encounters.

\bigskip
{\em Impact rate}
\medskip

Early analyses of the 
impact rate of dust particles measured with the Galileo dust 
detector during approaches to the Galilean moons 
showed a sharp peak within about half an hour of closest approach to the
moon (\cf\ {\it Gr\"un et al.,} 1997, 1998; {\it Kr\"uger et al.,} 1999b).
This indicated the existence of concentrations of dust particles
within a few satellite radii above the satellites' surfaces. 

\begin{figure}
\epsfxsize=12.5cm
\epsfbox{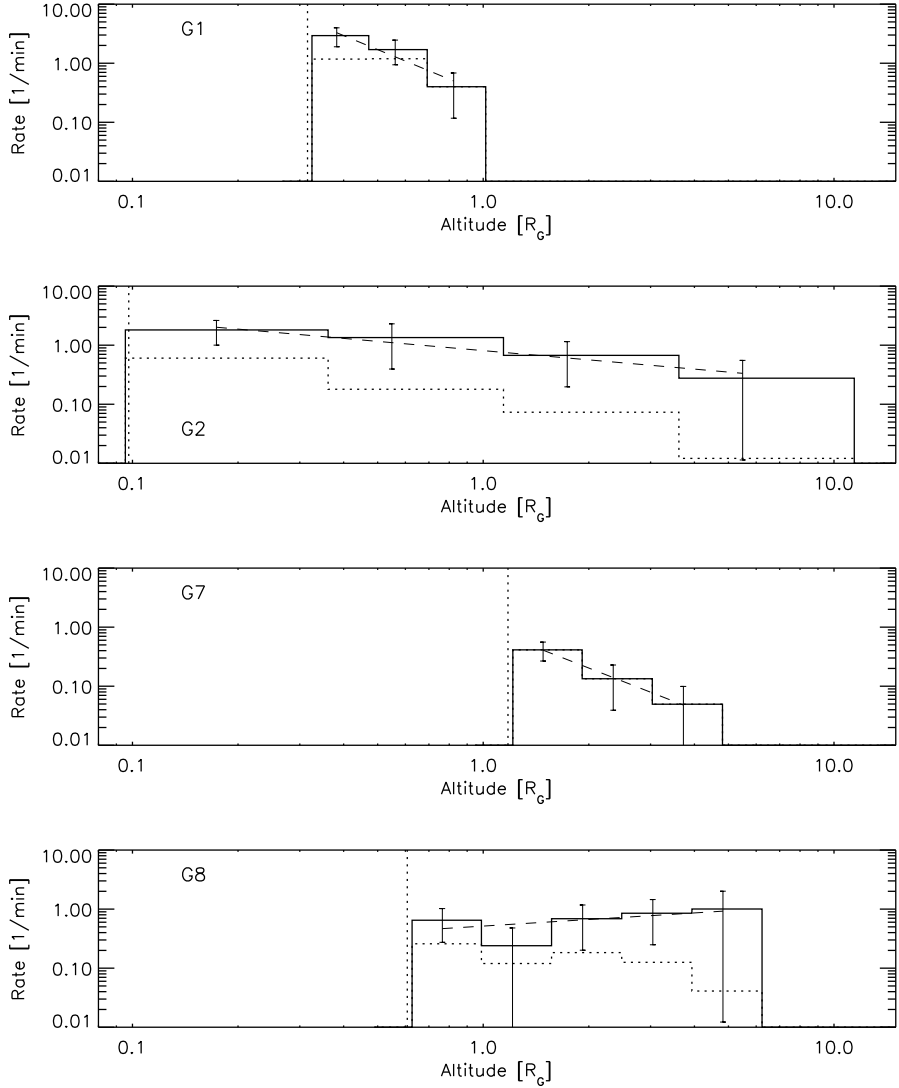}
\vspace{-4cm}
        \caption{\label{rate}
Impact rate of dust particles detected during the four Ganymede encounters.
The dotted histogram bins 
show the impact rate derived from the number of particles for which
their complete information has been transmitted to Earth. 
The solid histograms show the same rates, but corrected for incomplete data 
transmission. 
The vertical dotted lines indicate the minimum altitude reached by Galileo
at closest approach.
Error bars denote the $\sqrt{n}$ statistical uncertainty, with $n$ being
the number of particles for which the complete information has been 
transmitted. For G8 only particles with a calibrated velocity below 
$\rm 10\,\km\second^{-1}$ have been used, and the rate has been truncated 
at $\rm 6\, R_G$. Dashed lines are power law fits to the corrected impact 
rate (\cf\ Tab.~1).
}
\end{figure}

With the data set for Ganymede particles detected during all four 
encounters we can now calculate the impact rate 
of dust grains in the close vicinity of the moon (Fig.~\ref{rate}). 
To obtain the impact rate we define distance bins equally 
spaced on a logarithmic scale and divide the number of 
particles for which the complete set of parameters has been 
transmitted to Earth in a given distance bin by the time Galileo 
has spent in this bin (dotted lines). 

\bigskip
\centerline{\fbox{\bf Insert Figure \ref{rate}}}
\medskip

To correct for 
incomplete data transmission, we then multiply bin by bin the impact 
rate with the ratio between the number of counted particles
and the number of particles for which the complete 
data set has been transmitted. The impact rates corrected for
incomplete data transmission are shown as solid lines. In cases where 
only a solid line is visible no correction was applied because
the complete information of all particles in that bin is available
(G1 and G7). 

During the first three encounters (G1, G2 and G7)
the impact rate increases towards Ganymede.
Note specifically that for G7 no correction for incomplete 
transmission is needed because the full data set for all detected 
particles has been transmitted. Also, G1 needs only a 
correction in the innermost bins which steepens the slope slightly. 
Thus, G1 and G7 probably give the most reliable slopes for the spatial 
distribution of the grains in the dust cloud. Both fits give 
power law slopes steeper than $-2$.

For the fourth encounter (G8) no concentration of particles towards 
Ganymede is seen after correction for incomplete transmission. Note
that a slight concentration is obvious in the uncorrected rate
(dotted curve). This indicates that the slopes for the corrected  
curves have to be taken with some caution. 
Interestingly, the flattest slopes have been found for the encounters 
with the largest corrections for incomplete transmission 
(G2 and G8). On the other hand, the G2 flyby was the one which 
occurred closest to the north pole of Ganymede 
($\rm 80^{\circ}$ latitude).
We do not investigate variations of the slopes between 
individual encounters because of the large statistical uncertainties.
Spatial variations with respect to the flyby position relative to the
satellite will be addressed in a future investigation 
which will include data from Galileo's  Europa and Callisto flybys. 

\bigskip
{\em Mass distribution}
\medskip

\begin{figure}
\epsfxsize=12.5cm
\epsfbox{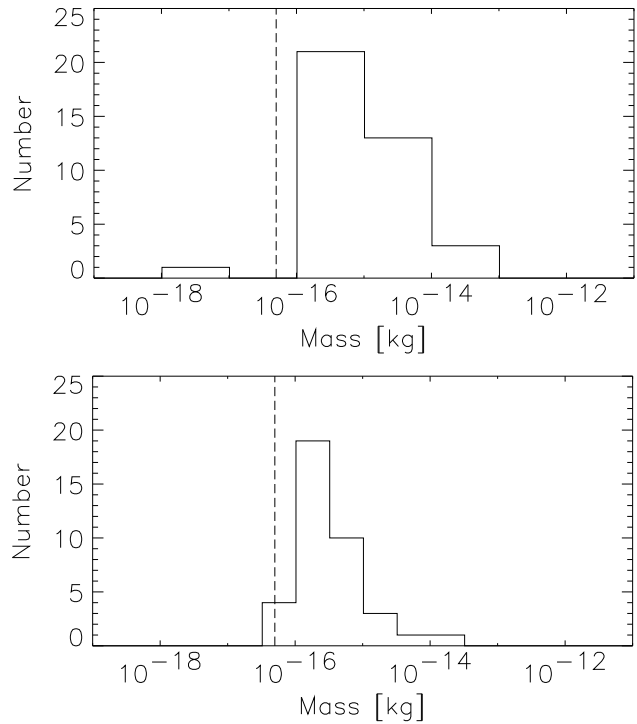}
\vspace{-4cm}
        \caption{
\label{mass_hist}
Mass distribution of the particles detected during the four Ganymede flybys.
The upper panel shows the distribution obtained by using the 
measured impact velocities derived from 
the instrument calibration. In the lower panel the velocity 
of Galileo relative to Ganymede has been assumed as the impact 
velocity in order to calculate the mass of the particle.
The dashed lines indicate the detection threshold for 
particles which approach the detector with the velocity of
Galileo relative to Ganymede (about $\rm 8 \km\second^{-1}$). Only the 38 particles 
with a velocity error factor $\rm VEF < 6$ ({\it Gr\"un et al.,}
1995) have been considered.
}
\end{figure}

The charge released by an impact of a dust particle onto
the target depends on the mass and the velocity of the 
grain ({\it Gr\"un et al.,} 1995).  Thus, in order to calculate 
the particle mass from the measured impact parameters 
one has to know the impact velocity. 
In Fig.~\ref{mass_hist} we show the mass distribution of the
particles from all four Ganymede flybys for which the velocity
could be reliably determined ($\rm VEF < 6$; 38 particles in total).

\bigskip
\centerline{\fbox{\bf Insert Figure \ref{mass_hist}}}
\medskip

In the upper panel of 
Fig.~\ref{mass_hist} the impact velocities derived from the
instrument calibration have been used to obtain 
the particle mass.  With this method the uncertainty of the 
impact velocity is typically a factor of 2 and that of the 
mass is a factor of 10.

The dust detector has a velocity-dependent
detection threshold ({\it Gr\"un et al.,} 1995). The threshold
for particles approaching with $\rm 8 \km\second^{-1}$
is shown as a dashed line. The mass distribution is incomplete
around and below this value.

The mass distribution is also affected by the low data transmission 
capability of Galileo and the data storage scheme in the instrument 
memory. As a result, nearly all particles lost in G1, G2 and G8 are 
in the mass range below  $\rm 10^{-15}\,kg$. If we assume 
that the lost particles are equally distributed over the mass bins
below this value, the maximum of the mass distribution is 
artificially too low by less than a factor of 2.

If the individual impact velocities were known with a higher 
accuracy than the 
typical factor of 2 uncertainty from the instrument calibration, the 
uncertainty in the mass determination could be reduced. 
Because the mean values of the measured impact velocities are in good 
agreement with the velocities of Galileo relative to Ganymede 
for the individual encounters (Tab.~1), we assume 
the latter ones as the particles' impact velocities and 
re-calculate the particle mass. This method has also 
successfully been applied to the mass distribution of 
interstellar dust 
particles ({\it Landgraf et al.,} 2000). 
The lower panel of Fig.~\ref{mass_hist}
shows the result for the Ganymede particles.
The width of the mass distribution is significantly smaller than 
that derived from the calibrated impact velocities.

\begin{figure}
\vspace*{-4cm}
\epsfxsize=10cm
\epsfbox{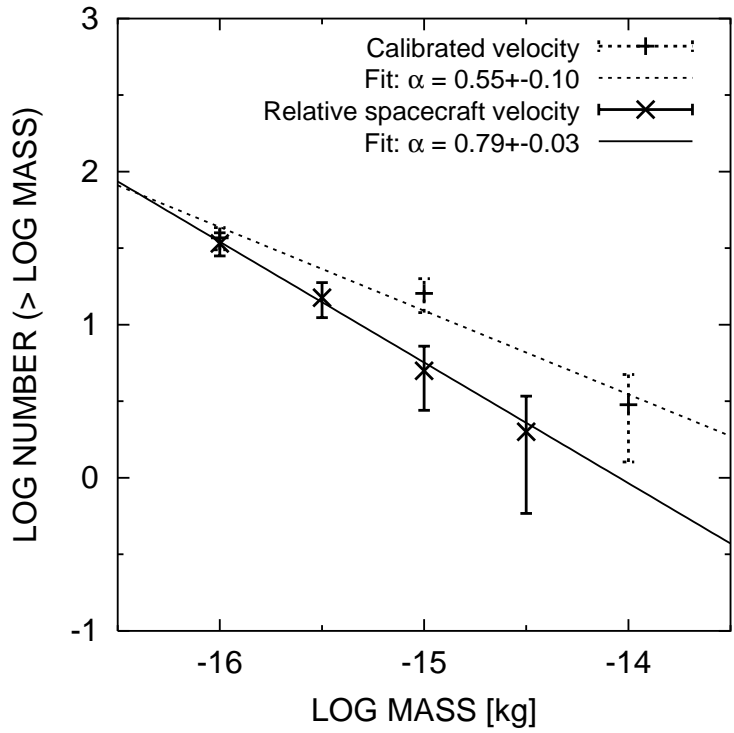}
        \caption{
\label{mass_bw}
Cumulative mass distributions of Fig.~\ref{mass_hist} and linear fits to 
the data. Vertical bars indicate the $\sqrt{n}$ statistical uncertainty. 
}
\end{figure}

The mass distribution of the grains allows for a simple check 
for the compatibility of the data with the hypothesis
of the impact origin of the detected particles.
We took the mass distributions shown in 
Fig.~\ref{mass_hist} and show linear fits to the cumulative 
distributions in Fig.~\ref{mass_bw}.
We only considered grains, the masses of which were definitely
well above the detector threshold ($M > 10^{-16}\kg$).
The slopes are in good agreement with the typical slopes 
one expects for
the impact ejecta ($0.5 \la \alpha \la 1.0$;
see, \eg, {\it Krivov and Jurewicz,} 1999)\nocite{krivov-jurewicz-1998}.

\bigskip
\centerline{\fbox{\bf Insert Figure \ref{mass_bw}}}
\medskip


\section{Model for the Impact Ejecta from Ganymede}

Detections of dust near Ganymede, and our results of the data processing
which reveal the concentration of dust towards Ganymede,
give rise to the following question: what is the physical mechanism
that produces the dust material?
A possibility of capturing the grains by the gravity of Ganymede is
dynamically ruled out.
In the 3-body problem ``dust grain -- Ganymede -- Jupiter'',
a close encounter of a grain with Ganymede may
convert a hyperbolic orbit with respect to Jupiter to an elliptical one,
but still bound to Jupiter and not to Ganymede.
Under some conditions, it is possible that interplanetary dust
grains are trapped in Jovian magnetosphere by electromagnetic forces 
({\it Colwell et al.,} 1998; {\it Thiessenhusen et al.,} 2000); 
but a strong concentration of dust towards Ganymede requires additional
mechanisms connected with Ganymede itself.
The recently discovered own magnetic field of Ganymede 
({\it Kivelson et al.,} 1998) \nocite{kivelson-et-al-1998}
seems to be too weak to do the job.
One could claim that geysers or volcanoes loft dust off Ganymede, but this
is very unconvincing and unsupported by the observations which show no
activity of this kind.
One could argue for a self-sustained distribution of dust which was
suggested to take place in Saturn's E ring
({\it Hamilton and Burns}, 1994),\nocite{hamilton-burns-1994}
but the strong gravity of Ganymede makes that improbable.
A conceivable guess one could make about the origin
of the Ganymede grains is their production via continuous bombardment
by interplanetary micrometeoroids.

In what follows, we construct a model to predict the density of the dust
cloud of Ganymede, produced by continuous hypervelocity impacts of
interplanetary projectiles onto the moon's surface.
We start with the main simplifying assumptions.
Two populations of the debris particles are considered:
those which move on ballistic trajectories and therefore fall back to the
satellite shortly after they are ejected,
as well as  those which are fast enough to escape from Ganymede into
circumjovian space.
We assume that these ejecta move in Keplerian trajectories~--- 
pieces of ellipses and hyperbolas, respectively.
This is justified by estimates of the perturbing forces.
The tidal gravity of Jupiter is only
important outside the Hill sphere ($R_H \approx 12 \rm \,R_G$).
Next, for the grains with masses greater than  $10^{-16}\kg$,
to which we confine our analysis, the radiation pressure force
is less than 10\% of Ganymede's gravity even at $\rm 10 \,R_G$ from the
moon. The Coriolis force is comparable to the satellite gravity at
$\rm \sim 5\, R_G$, but it cannot change appreciably the radial
distribution of the ejecta.
The only force that can possibly make the motion essentially
non-Keplerian is the Lorentz force. Still this force, which is very
important for $10^{-16}\kg$ grains moving about Jupiter
({\it Colwell et al.,} 1998; {\it Krivov et al.,} 2000),
can probably be neglected for short-living grains in the vicinity of
Ganymede.
Assuming a strength of the Jovian magnetic field at Ganymede of $\sim
10^2\nT$, 
the velocity of the grains
to be equal to the Ganymede velocity relative to the corotating
jovian magnetic field ($180\km\second^{-1}$),
and electrostatic potentials of grains
of $+3\V$ ({\it Hor\'anyi et al.,} 1993),
we find that at $5R_G$ the Lorentz force amounts to $\sim 30$\%
of the satellite gravity. 
The intrinsic magnetic field of Ganymede is ten times stronger 
($\la 10^3\nT$; {\it Kivelson et al.,} 1998),\nocite{kivelson-et-al-1998}
but the velocity of the grains relative to Ganymede and its magnetic field
is only of the order of $1\km\second^{-1}$,
which means that the relative strength of the Lorentz force caused
by the satellite's magnetic field is one order of magnitude weaker.
We conclude, especially taking into account the uncertainty in
grains' charges, that under some conditions the Lorentz
force might lead to noticeable effects.
However, incorporation of the electromagnetic effects into the
model would require an extensive dedicated modelling, including
implementation of models of Ganymede's magnetosphere and performing
the charging calculations. This would make little sense in view of the
scarcity of the Galileo dust data. Thus this task is beyond the scope of
the present paper.

We also assume that the circumsatellite dust cloud is in a steady state.
Finally, we neglect possible effects of
non-isotropy of the impactor flux
(\eg, {\it Colwell}, 1993)\nocite{colwell-1993}
and assume that the cloud is spherically symmetric.
Thus, in this paper we do not consider variations in the spatial density 
from flyby to flyby which may be caused by spatial or temporal 
variations of the dust cloud surrounding Ganymede. 
Again, such an investigation
is not possible with the present data set of 45 ejecta grains. 

\bigskip
{\em Impactor flux and ejecta yield}
\medskip

\begin{figure}
\vspace*{-4cm}
\epsfxsize=11cm
\epsfbox{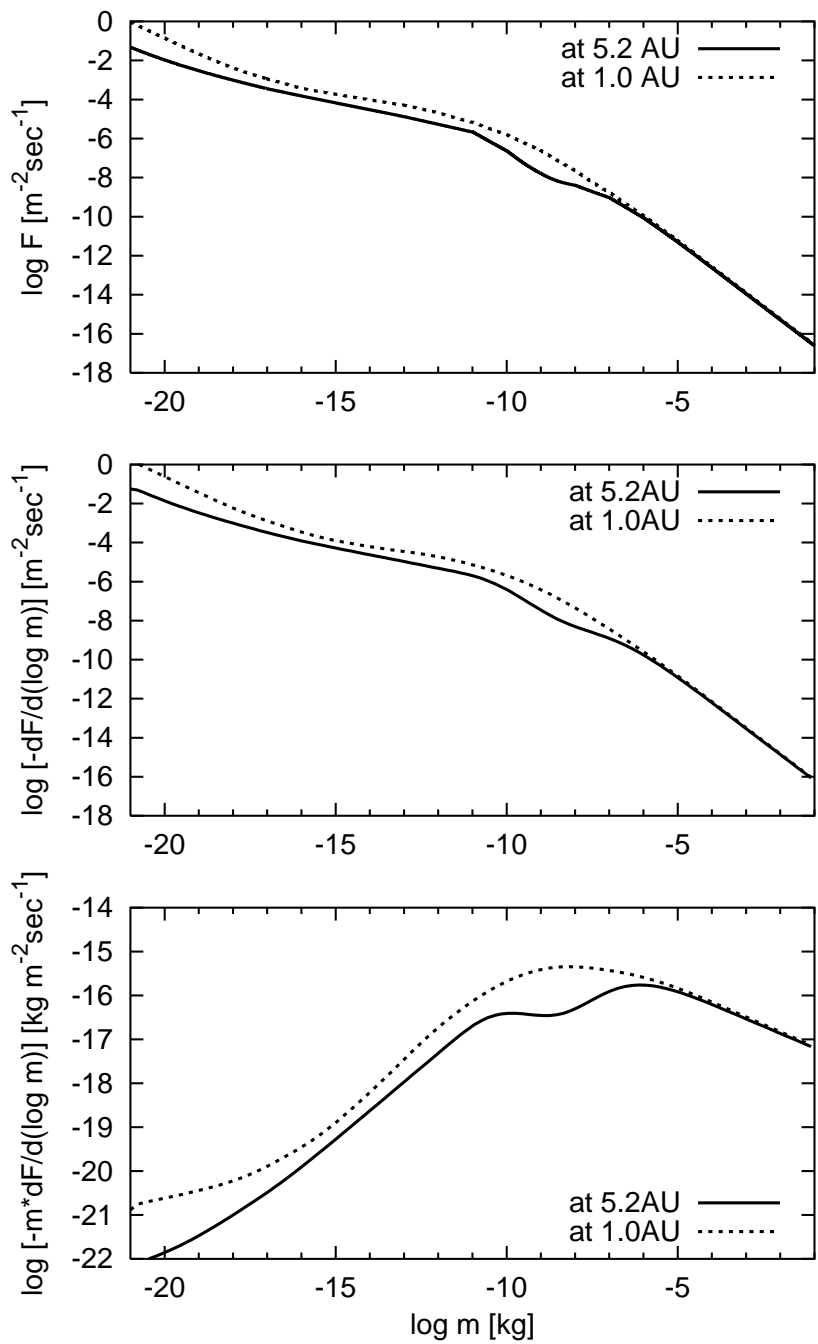}
        \caption{
\label{divine}
Fluxes of interplanetary grains onto a sphere with unit cross section
($\pi R^2 = 1$), moving around the Sun in a circular Keplerian orbit
with radius of $5.2\AU$ (solid lines) and $1\AU$ (dashed lines),
according to Divine's (1993) model. 
Top: the cumulative flux;
middle: the differential flux per unit log mass interval;
bottom: the differential mass flux per unit log mass interval.
}
\end{figure}

To predict the flux of the interplanetary impactors, we use 
{\it Divine}'s (1993) model. \nocite{divine-1993} We have calculated the
fluxes of interplanetary grains onto a sphere with unit cross section
($\pi R^2 = 1$), moving around the Sun in a circular Keplerian orbit
with a radius of $5.2\AU$ (heliocentric distance of Jupiter).
Figure~\ref{divine} shows, as a function of particle's mass,
the cumulative flux onto this sphere, the differential flux per mass
decade, as well as the differential mass flux per mass decade.
For comparison, the same fluxes at $1\rm AU$ are also shown.
The total mass flux at $\rm 5.2AU$ ``far'' from Jupiter
(\ie, before its gravity is taken into account) is
$F_\infty = 8.4 \times 10^{-16}\kg\m^2\second^{-1}$.
It is dominated by grains with masses $\sim 10^{-8}\kg$
(or with sizes $\sim 100\mum$).

\bigskip
\centerline{\fbox{\bf Insert Figure \ref{divine}}}
\medskip

The main contribution to the flux and mass flux is made by the ``halo
population'' ({\it Divine,} 1993), composed of particles in orbits with
moderate eccentricities, but randomly distributed inclinations
between $0^\circ$ and $180^\circ$ (thus including grains in both prograde
and retrograde orbits). The velocities of interplanetary
grains with respect to the Sun are very high:
$\approx 20\km\second^{-1}$.
Their mean velocity relative to Jupiter, because of the random
distribution of inclinations, is nearly the same,
$\approx 20\km\second^{-1}$.
Obviously, the dispersion is also high:
since the orbital velocity of Jupiter round the Sun is
$\approx 13\km\second^{-1}$, the velocities of individual particles
may range from $7$ to $33\km\second^{-1}$.
For comparison, {\it Colwell and Hor\'anyi} (1996)
\nocite{colwell-horanyi-1996} adopt for this ``Oort cloud population''
of grains the velocity range of 8 to $34\km\second^{-1}$ with 
a mean of $23.6\km\second^{-1}$.

We have to take into account the gravitational focussing
by Jupiter, which increases the interplanetary flux $F$
at Ganymede ($r = 15 R_J$):
(i)~the speed of the grains $v$ becomes larger than that
far from the planet, $v_\infty$, and
(ii)~the spatial density of dust $n$ gets larger than
the one far from Jupiter, $n_\infty$.
Assuming $v_\infty = 20 \km\second^{-1}$ and applying the energy
integral, we find $v / v_\infty = 1.3$ and $v \approx 25\km\second^{-1}$.
The velocity $v$ has the meaning of the mean velocity of impactors
with respect to {\em Jupiter} at Ganymede's distance.
Nevertheless, since the incoming directions of the projectiles are
broadly distributed, we can roughly take this value as the mean projectile
velocity with respect to {\em Ganymede}.
We note, however, that the velocities of individual grains striking 
Ganymede may range from nearly zero to about $50\km\second^{-1}$.
The former takes place when Ganymede is at opposition to the Sun for the
prograde particles hitting its trailing hemisphere; the latter is attained
at the same moment for retrograde particles impacting the leading hemisphere.
Next, using the formulae by
{\it Colombo et al.} (1966)\nocite{colombo-et-al-1966},
we find $n / n_\infty = 1.4$. Therefore,
\be \label{focussing}
 {F \over F_\infty} = 
 {v \over v_\infty} {n \over n_\infty} 
 = 1.8,
\ee
so that the mass flux at Ganymede is
$F = 1.5 \times 10^{-15}\kg\m^2\second^{-1}$.

We consider the impact ejecta production now.
The efficiency of the material ejection in a cratering event is
characterised by $Y$, the characteristic yield, defined as the ratio of
the ejected mass to the projectile mass.
For hypervelocity impacts into ices, the typical yields $Y$ were
reported to range from $\sim 10^3$ to $10^6$
(\eg, {\it Lange and Ahrens,} 1987;\nocite{lange-ahrens-1987}
{\it Koschny, priv. comm.}).
From ({\it Koschny, priv. comm.}),
for pure ice, for $10^{-8}\kg$ impactors,
and for an impact speed of $25\km\second^{-1}$,
the yield is $Y \approx 1 \times 10^4$.
The mass production rate from Ganymede's surface is then
\be
  M^+ = F Y S = 330 \kg\second^{-1},
\ee
where $S = \pi R_G^2 = 2.2 \times 10^{13}\m^2$ is the cross section area
of Ganymede.

The mass distribution of the ejecta is commonly represented by a power
law: $N^+(>M) \propto M^{-\alpha}$, where $M$ is the grain mass
and $N^+(>M)$ is the number of particles with masses $> M$ ejected from
the moon per second.
The distribution is normalised to the total amount of the ejecta $M^+$:
\be \label{N+}
 N^+(>M) = {1-\alpha \over \alpha} {M^+ \over M_{max}}
 \left( M_{max} \over M \right)^{\alpha},
\ee
where $M^+$ is the largest mass of the ejecta.
A plausible slope of the ejecta mass distribution
is $\alpha \approx 0.83$
(e.g., {\it Koschny, priv. comm.}; \nocite{koschny-gruen-1996b}  
{\it Krivov and Jurewicz,} 1999)\nocite{krivov-jurewicz-1998}
which is consistent with the exponent found from the
Galileo data (Fig.~\ref{mass_bw}).
For the heaviest ejecta fragment, we take $M_{max} = 10^{-8}\kg$,
which is close to the typical mass of the impactors
(the dependence of this parameter is weak though).
In what follows, we always calculate $N^+ \equiv
N^+(> 10^{-16}\kg)$, which correspond to the grains  ``safely'' above the
detection threshold, \cf\ Fig.~\ref{mass_hist}).

\bigskip
{\em Ejecta speed distribution}
\medskip

In this study, we assume no dependence between the speeds of the
ejected grains and their masses, because laboratory experiments
still do not evidence any strong correlation between the two
quantities
(\eg,
{\it Nakamura and Fujiwara}, 1991)\nocite{nakamura-fujiwara-1991}.
Thus, regardless of how massive the ejecta are, we
take the ejecta speed distribution in the form
(see, \eg, {\it St\"offler et al.}, 1975,\nocite{stoeffler-et-al-1975}
{\it Hartmann}, 1985)\nocite{hartmann-1985}
\be\label{Psi}
  \Psi(>u) = (u/u_0)^{-\gamma},
\ee
where $\Psi(>u)$ is the fraction of the material ejected at speeds
$> u$.

This adds two parameters to the model:
the ``lower cut-off velocity'' $u_0$ and the distribution slope $\gamma$.
The values of these parameters were determined in impact experiments.
A plausible range for $u_0$ could be from a few $\m\second^{-1}$ to
several hundred $\m\second^{-1}$
(\eg, {\it St\"offler et al.,} 1975;\nocite{stoeffler-et-al-1975}
{\it Hartmann,} 1985)\nocite{hartmann-1985}.
The slope of the speed distribution $\gamma$ may range from 1
to somewhat greater than 2
(\eg, {\it Frisch,} 1992;\nocite{frisch-1992}
see also a discussion in
{\it Colwell and Esposito}, 1993).\nocite{colwell-esposito-1993}

A remark should be made about the directions of the ejecta
velocities.
From the laboratory experiments it is known that even oblique
impacts lead to ejection of the surface material into a cone normal
to a target surface
({\it Nakamura and Fujiwara,} 1991)\nocite{nakamura-fujiwara-1991}.
By numerical experiments, we have checked that the spatial distribution 
of the ejecta depends only weakly on the opening angle of this cone.
Also, laboratory measurements suggest that most of the debris are ejected
at high angles to the surface~--- such as $60^\circ$
({\it Koschny, priv. comm.};
{\it Burchell \etal,} 1998;\nocite{burchell-et-al-1998}
also {\it M. J. Burchell}, priv. comm.)
or even $80^\circ$ ({\it Frisch,} 1992).\nocite{frisch-1992}
Besides, some authors observed vertical plumes of fast ejecta
({\it Hartmann,} 1985).\nocite{hartmann-1985}
For these reasons, and to alleviate the analytical derivations,
we assume for simplicity that the trajectories of the ejecta are straight
lines normal to the surface of Ganymede (``degenerate Keplerian
ellipses'').

\bigskip
{\em Energy constraints on the model parameters}
\medskip

The model as  a whole contains five parameters.
Three parameters come from the ejecta mass distribution
(characteristic yield $Y$,
slope $\alpha$,
and the maximum mass of a fragment $M_{max}$),
while two others pertain to the ejecta speed distribution
(the lower cut-off velocity $u_0$ and the slope $\gamma$).
The dependence of the results on $\alpha$ and $M_{max}$ is only
moderate, and besides, changes in their values can always be absorbed
in a value of the characteristic yield.
This leaves $Y$, $u_0$, and $\gamma$ as the three most important
model parameters.
Of course, the parameters describing the projectile flux, most notably the
mass flux of the impactors $F$, are implicit in the results.

Some constraints on the parameters $Y$, $u_0$, and $\gamma$ can be placed
by considering the energy balance.
The ratio of the kinetic energy carried by the ejecta cloud, $K_e$,
to the impactor's kinetic energy, $K_i$, must be less than unity,
because part of $K_i$ is spent for comminution and heating.
The ratios determined in the hypervelocity impact experiments vary from
$\sim 0.2$ to $\sim 0.5$
(\eg, {\it Asada,} 1985;\nocite{asada-1985}
{\it Hartmann,} 1985)\nocite{hartmann-1985}.
We assume $f \equiv K_e/K_i = 0.3$.

\begin{figure}
\epsfxsize=11cm
\epsfbox{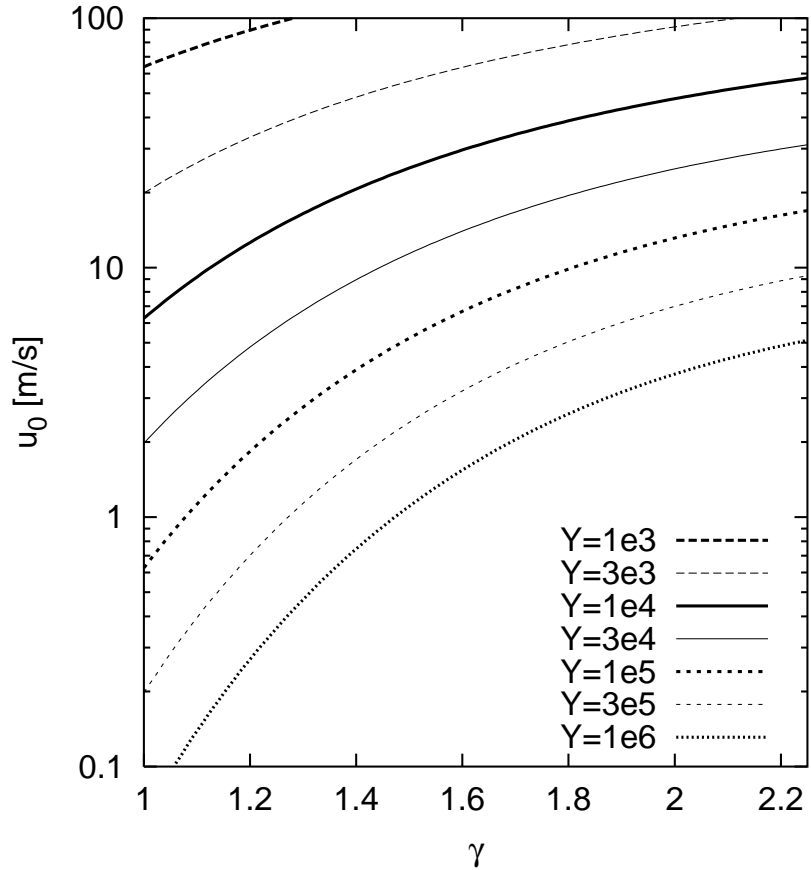}
        \caption{
\label{energy}
Constraints on the model parameters $Y$, $u_0$, and $\gamma$
coming from the consideration of the kinetic energy.
Shown, for several values of the characteristic yield $Y$, are the lines
of constant ratio $K_e/K_i =0.3$, where $K_i$ and $K_e$ are
respectively the kinetic energy of an impactor and the ejecta it produces.
The impact speed is taken to be $25\km\second^{-1}$.
The speeds of the ejecta are assumed to obey a power law (Eq.~\ref{Psi})
with a slope $\gamma$, a lower cut-off $u_0$ and the upper cut-off
$u_{max} = 3\km\second^{-1}$.
}
\end{figure}

Let us express $K_i$ and $K_e$ through the parameters of the model.
Obviously, $K_i = m v^2 /2$, and
\be
  K_e = Ym \int_{u_0}^{u_{max}} {u^2 \over 2} \psi(u) du,
\ee
where $m$ is the impactor mass, $v$ is its speed,
$u_{max}$ is the speed of the fastest ejecta fragments,
and $\psi(u) \equiv -\Psi^\prime(>u)$
(hereafter prime means derivative) is the differential speed distribution,
with $\Psi(>u)$ being the cumulative distribution given by Eq.~(\ref{Psi}).
We have:
\be
  f = Y {\gamma \over 2-\gamma} \left( u_0 \over v \right)^2
      \left[ \left( u_0 \over u_{max} \right)^{\gamma -2} - 1 \right]
  \qquad
  (\gamma \ne 2) .
\ee
(If $\gamma = 2$, a logarithmic dependence results.)
In Fig.~\ref{energy}, the lines $f=0.3$ are plotted on the $\gamma, u_0$-plane
for several values of $Y$. For $Y=10^4$, which we have chosen as the most
plausible value, and for $\gamma = 1.2$, 1.6, and 2.0, one gets
$u_0 = 13$, 30, and $48\m\second^{-1}$, respectively.
In these calculations, we assumed $u_{max} = 3\km\second^{-1}$.
The results do not depend on this speed severely.
For example, with $u_{max} = 4\km\second^{-1}$, for the same $Y$ and
$\gamma$ one gets $u_0 = 10$, 27, and $45\m\second^{-1}$.
The values of $u_0$ that we derived are consistent with impact experiment
data (\eg, {\it St\"offler et al.,} 1975;\nocite{stoeffler-et-al-1975}
{\it Hartmann,} 1985).\nocite{hartmann-1985}

\bigskip
\centerline{\fbox{\bf Insert Figure \ref{energy}}}
\medskip

\bigskip
{\em Number density of dust}
\medskip

Given a dust production rate $N^+$,
which can be calculated using
Eq.~(\ref{N+}),
and assuming the ejecta speed distribution (Eq.~\ref{Psi}),
we compute now the steady-state number density of the ejecta, $n$,
as a function of $r$, the distance from the centre of Ganymede.

Consider first the ejecta in {\em bound} orbits, \ie, the grains
ejected at speeds $u < u_{esc}$, where $u_{esc} = 2,750 \m\second^{-1}$ is
the escape velocity from Ganymede's surface.
Remember that we assume the grains to move in rectilinear Keplerian
trajectories.
As follows from Eq.~(\ref{Psi}) and from the energy integral of the
two-body problem, the fraction of ejecta, ejected in orbits with semimajor
axes $>a$, is
\be\label{intermed1}
  \Psi(>a) = (u_0/u_{esc})^\gamma (1 - R_G/(2a))^{-\gamma/2}.
\ee
The steady-state number of grains with semimajor axes $(a, a+da)$
at distances $(r, r+dr)$ from Ganymede's centre, where $r \le 2a$, equals
\be\label{intermed2}
  - 2 N^+ \Psi^\prime(>a)da dt,
\ee
where the factor 2 appears because a grain reaches the distance $r$
twice~--- moving from and back to the moon,
and $dt$ is the time interval it takes for a grain to move from $r$
to $r+dr$.
Dividing Eq.~(\ref{intermed2}) by the volume $4 \pi r^2 dr$ of a
spherical layer $(r, r+dr)$ and integrating over semimajor axes of grains
that reach distance $r$, one finds the steady-state number density of
grains at distance $r$:
\be\label{intermed3}
  n_{bound}(r) = {N^+ \over 2\pi r^2}
    \int_{r/2}^{+\infty}(-\Psi^\prime(>a)) {dt \over dr} da.
\ee
From the energy integral,
\be\label{intermed4}
  {dt \over dr} = { 1 \over u_{esc} } \sqrt{r/R_G \over 1 - r/(2a)}.
\ee
Inserting Eq.~(\ref{intermed1}) and Eq.~(\ref{intermed4}) into
Eq.~(\ref{intermed3}), and using a new integration variable $x = 2a/r$,
after some algebra we find
\be \label{n bound}
  n_{bound}(r) =
  {N^+ \over 4 \pi r^2}
  {\gamma \over u_{esc}}
  \left(
    u_0 \over u_{esc}
  \right)^\gamma
  \sqrt{R_G \over r} \\ 
   \cdot 
  \int\limits_1^{+\infty}
    {dx
     \over
     \left(
         1 - R_G r^{-1} x^{-1}
     \right)^{\gamma/2 + 1}
     x^2 \sqrt{1 - x^{-1}}
    } .
\ee

Now we perform a similar derivation for the
{\em escaping} grains with velocities $u \ge u_{esc}$ at the surface.
The steady-state number of grains ejected at speeds $(u, u+du)$
($u \ge u_{esc}$) at distances $(r, r+dr)$ from Ganymede's centre is
\be\label{intermed2a}
  - N^+ \Psi^\prime(>u)du dt,
\ee
where $dt$ is again the time that a grain needs to move from $r$ to
$r+dr$.
Dividing Eq.~(\ref{intermed2a}) by the volume of a
spherical layer $(r, r+dr)$ and integrating over $u \ge u_{esc}$ there
results:
\be\label{intermed3a}
  n_{unbound}(r) = {N^+ \over 4\pi r^2}
    \int_{u_{esc}}^{+\infty}(-\Psi^\prime(>u)) {dt \over dr} du.
\ee
Here,
\be\label{intermed4a}
  {dt \over dr} = { 1 \over 
  u_{esc}  \sqrt{ (u/u_{esc})^2 - 1 + R_G/r} } .
\ee
Substituting Eq.~(\ref{Psi}) and Eq.~(\ref{intermed4a}) into
Eq.~(\ref{intermed3a}), and using an integration variable
$x = u/u_{esc}$, after some transformations one obtains
\be \label{n unbound}
  n_{unbound}(r) =
  {N^+ \over 4 \pi r^2}
  {\gamma \over u_{esc}}
  \left(
    u_0 \over u_{esc}
  \right)^\gamma \\
   \cdot 
  \int\limits_1^{+\infty}
    {dx \over
     x^{1 + \gamma}
     \sqrt{x^2 - 1 + R_G/r}
    } .
\ee

\begin{figure}
\vspace*{-4cm}
\epsfxsize=11cm
\epsfbox{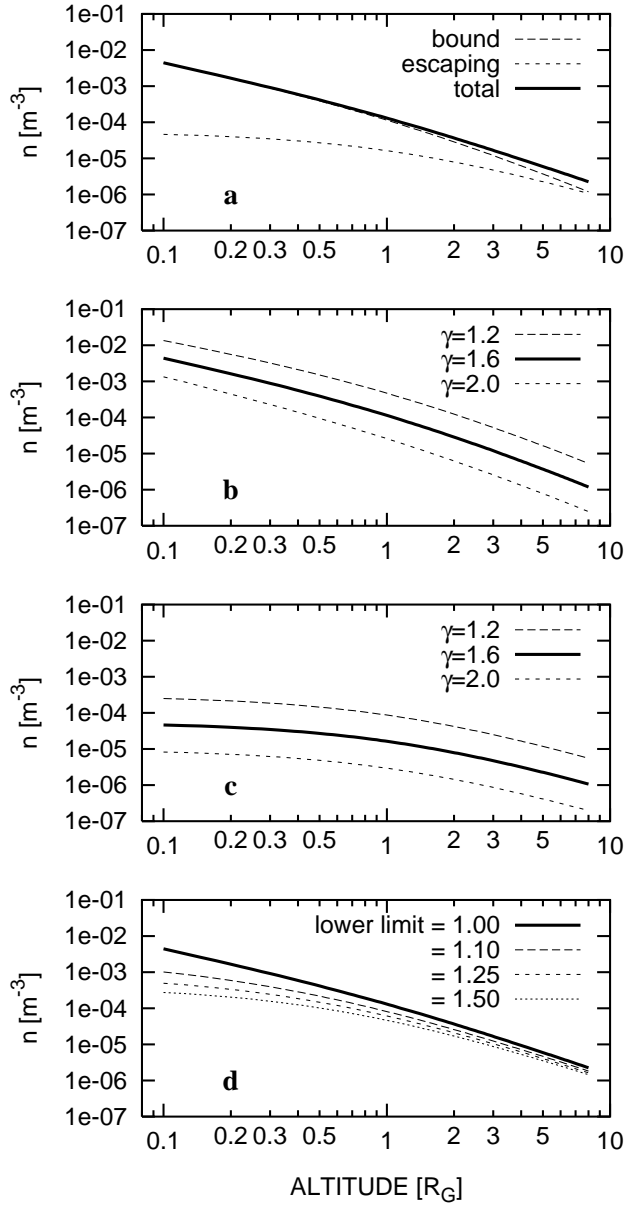}
\vspace*{2cm}
        \caption{
\label{parms}
Plots of the functions $n_{bound}(r)$ (Eq.~\ref{n bound})
and $n_{unbound}(r)$ (Eq.~\ref{n unbound}).
Adopted values of the parameters:
$Y=1 \times 10^4$, $u_0 = 30\m\second^{-1}$.
(a) These two functions and their sum for a fixed $\gamma = 1.6$.
(b) Number density of grains in bound orbits for several values of
$\gamma$.
(c) The same, but for escaping grains.
(d) Number density of dust (in bound and unbound orbits together)
for a fixed $\gamma = 1.6$, but for different
lower integration limits in Eq.~(\ref{n bound}),
which simulates oblique ejections of grains from the surface.
See text for further explanations.
}
\end{figure}

Let us discuss some properties of the functions
$n_{bound}(r)$ (Eq.~\ref{n bound})
and $n_{unbound}(r)$ (Eq.~\ref{n unbound}).
To get the actual number density at a distance $r$, we 
sum up the two contributions:
$n(r) = n_{bound}(r)+  n_{unbound}(r)$.
As seen from Eqs.~(\ref{n bound}) and (\ref{n unbound}),
$n_{bound}(r)$ decreases faster with increasing $r$
than $n_{unbound}(r)$, so that
the relative contribution of escaping grains increases with distance.
However, how large is this contribution at a
given distance from the satellite?
For one set of the model parameters, 
namely $Y=1 \times 10^4$ and $u_0 = 30\m\second^{-1}$,
we made calculations by using 
both Eqs.~(\ref{n bound}) and (\ref{n unbound}) and found that
the number densities of the escaping grains and the particles
falling back to the moon become comparable at the altitude of 
$\rm \sim 8 \,R_G$
from Ganymede (Fig.~\ref{parms}a).
As the Galileo detections have been made closer to the moon,
we can conclude that the majority of the ejecta particles identified by 
the dust instrument, if they had not been caught by the detector,
eventually would have fallen back onto Ganymede.

Somewhat counterintuitively, the radial slope of the number density
is not very sensitive to $\gamma$.
For all reasonable values of the ejecta distribution slope
$\gamma$, the number density is close to a power law 
$n(r) \propto r^{-\nu}$ with $\nu \sim 2$ to $3$
(see Fig.~\ref{parms}b for grains in bound orbits and
Fig.~\ref{parms}c for escaping grains).

Remember that, in derivation of
Eqs.~(\ref{n bound}) and (\ref{n unbound}), we assumed the ejecta to move
radially from and to the surface of Ganymede.
This assumption is partly reflected by the lower integration limit of 1
in Eq.~(\ref{n bound}). We calculated the integral with integration limits
greater than unity, to simulate the particles leaving the surface not
vertically. The results (Fig.~\ref{parms}d) show that at small altitudes
the number density gets smaller as compared to the case of vertical
ejections, and the curves slope more gently.

\bigskip
\centerline{\fbox{\bf Insert Figure \ref{parms}}}
\medskip


\section{Comparison of the Model with Galileo Data}

Comparison of the available data
with theory is accomplished by constructing the dependencies
of the frequency of the detections as a function of radial
distance from the moon.
To reach this goal, we use the impact rates depicted by solid
lines in Fig.~\ref{rate}. 
Dividing the rates by the effective 
spin-averaged detector area,
we obtain the fluxes $(\m^{-2}\second^{-1})$.
Then we divide the results by the mean impact velocity
for a given flyby, which results in mean number densities
$(\m^{-3})$ in various distance bins explained in Sect.~2.

On the other hand, we use our model to predict the number density
as a function of distance. 
We used Eqs.~(\ref{N+}), (\ref{n bound}) and (\ref{n unbound}).
As discussed above, we chose the yield $Y = 10^4$ and took
parameters $u_0$, and $\gamma$ in the ranges compatible
with laboratory impact experiments, using the values
that satisfy the relation $K_e/K_i = 0.3$.
As regards two other, less important, model parameters, the values
$\alpha = 0.83$ and $M_{max} = 10^{-8}\kg$ 
(\cf\ {\it Asada,} 1985;\nocite{asada-1985}
{\it O'Keefe and Ahrens,} 1985;\nocite{okeefe-ahrens-1985}
{\it Koschny, priv. comm.}) 
were used.

\begin{figure}
\vspace*{-1.5cm}
\epsfxsize=12cm
\epsfbox{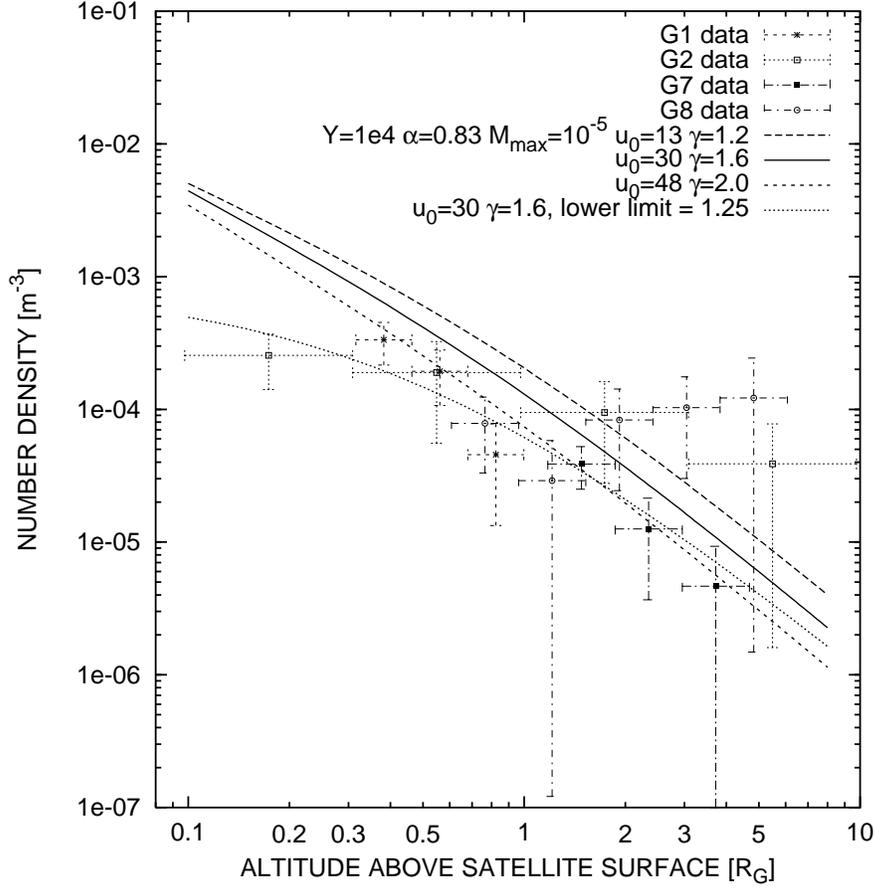}
        \caption{
\label{ganymede_bw}
The number density of dust as a function of distance from the centre
of Ganymede, derived from the data (symbols with error bars) and
predicted by the model (lines).
Horizontal bars for the data symbols indicate distance bins
which were used in processing the data (see text for details),
whereas vertical ones reflect $\rm \sqrt{n}$ errors due to a limited
number of impacts.
Theoretical curves are given for several plausible choices
of the model parameters, listed in the legend. 
For one of the models, we show the result for simulated non-vertical
ejections of dust from the surface (the lower limit in Eq.~(\ref{n bound})
was taken to be 1.25 instead of 1).
All the lines show the total number density of ejecta
(falling back and escaping) produced by interplanetary
impactors.
Here we do not consider variations in the spatial density 
from flyby to flyby which may be caused by spatial or temporal 
variations of the dust cloud surrounding Ganymede. 
}
\end{figure}

The results (both the data points and theoretical curves) are displayed
in Fig.~\ref{ganymede_bw}, showing a reasonably good agreement
between the data and the model.
For example, $Y = 10^4$, $u_0 \in [13, 48]\m\second^{-1}$, and $\gamma
\in [1.2, 2.0]$ are quite compatible with the data for G1 and G7,
which represent the most reliable number density slopes.
We cannot leave without notice, however, that 
the theoretical curves seem to be somewhat steeper than the
distribution of the data points for G2 and G8.
Either this effect is caused by statistical errors (see Section 2),
or it may be attributed to some simplifying assumptions of our model (see
Sections 3 and 5).
Note that using the lower integration limit in Eq.~(\ref{n bound})
larger than 1, which simulates the oblique ejections from the surface,
decreases the predicted number density at lower altitudes, resulting
in a better compatibility with some of the data points (\eg,
leftmost G2 point in Fig.~\ref{ganymede_bw}).

\bigskip
\centerline{\fbox{\bf Insert Figure \ref{ganymede_bw}}}
\medskip

It should be emphasised that in the present paper, we do not make
any systematic attempt to constrain the poorly known parameters of
the model from the data.
One reason for that is scarcity of the data.
Another one is that some of the parameters (such as $\alpha$,
$M_{max}$, and $\gamma$) affect the modelling results only slightly,
while the others (such as $Y$, $u_0$, and the parameters describing
the impactor flux) are strongly cross-correlated and cannot be
constrained independently.
A weak dependence of the predicted number density on some of the
parameters would certainly be a serious handicap, 
if our goal were to retrieve the poorly known parameters.
However, the same circumstance~--- a weak sensitivity of the model results
to the model parameters~--- turns to a serious advantage, if we are
merely checking the compatibility of the data with our model.
The fact that the model agrees with the data over a wide
range of the parameters strengthens our conclusion
about the likelihood of the impact origin of the detected grains.

In addition to the study of the debris produced by impacts of
interplanetary grains, we have also estimated a possible effect 
of interstellar grains as impactors.
The result 
is that their contribution is negligible.
The reason is that, although the flux of interstellar grains at the
heliocentric distance of Jupiter is known to exceed that
of the interplanetary particles, the {\em mass} flux in the considered
mass range is about 4 orders of magnitude smaller
({\it Landgraf et al.,} 2000).\nocite{landgraf-et-al-2000}
Hence the flux of the ejecta produced by interstellar grains is
also substantially smaller than the one caused by interplanetary projectiles.


\section{Conclusions and Discussion}

In this paper we examined the dust impacts registered by the Galileo
dust detector in the immediate vicinity of Ganymede
during four close flybys of this Jovian moon.
Analyzing impact directions and velocities and the mass distribution, 
as well as spatial locations of the events, we have evidenced
that the particles did originate from the moon.
In an attempt to find out the specific physical mechanism
that ejected the particles off the surface, we
checked whether these ``Ganymede dust grains'' are impact debris
produced by hypervelocity impacts and, 
which impactors are responsible for the ejections.
To do so, we have constructed a model for the impact ejecta from
Ganymede and compared the modelling results with the data.
The comparison made it clear that the data 
are fully compatible with 
the impact origin of the Ganymede dust grains.
Of course, this result cannot be treated as an evidence of the impact
origin of the detected dust.
With our modelling, we have only demonstrated that the data do not
contravene the impact scenario, given our current knowledge of
the physical conditions in the Jovian system as well as available
laboratory data of hypervelocity impacts.

If the impact scenario is true, then the class of projectiles
responsible for the formation of a circumsatellite cloud of dust grains,
several tens of which were sensed by the detector onboard Galileo,
is almost definitely interplanetary micrometeoroids.
The derived mass distribution of the detected grains, as well as
the impact rates actually measured by the dust instrument, are in fairly good
agreement with what we get from the model of hypervelocity impacts of
interplanetary dust particles (IDPs),
assuming contemporary models of IDP flux at a heliocentric distance
of Jupiter and a low-temperature ice target.
The interstellar grains should be much less
efficient as impactors in producing the collisional debris, because of
the much lower mass influx of interstellar dust onto the Ganymede surface
compared to that of the IDPs.

As follows from elementary estimates, the particles that reach
an altitude of one satellite radius must have a starting velocity
in excess of $2\km\second^{-1}$.
This confirms laboratory results that some of the
hypervelocity impact ejecta from icy targets attain very high speeds
({\it Frisch,} 1992).\nocite{frisch-1992}
Nevertheless, more than a half of the grains which could
be found (and have been actually detected by Galileo) at altitudes
less than about eight satellite radii above the surface, are not the
particles escaping into Jovian space.
They are slower ejecta, destined to fall back to Ganymede typically
within several minutes to several hours after ejection.
Such grains surround the satellite all the time as a result of
continuous bombardment of the surface by IDPs.
Further out, starting from distances of $\approx 8R_G$, the escaping
grains start to dominate the number density.

It seems useful to give some general estimates concerning the
mass budget of the dust cloud of Ganymede.
The mass flux of IDPs bombarding the satellite
surface is estimated to be $\sim 30\g\second^{-1}$
(dominated by IDPs with $m \sim 10^{-8}\kg$).
With the characteristic yields of $\sim 10^4$, 
we then estimate that 
as much as $\sim 10^2$ to $\rm 10^3\kg\second^{-1}$ of the moon's surface
material, is ejected into space.
Depending on the ejecta speed  distribution adopted, the mean
lifetime of the ejecta ranges from tens of seconds to
several hundred seconds.
Note that these values are dominated by the
slowest ejecta~--- for the grains that reach the altitudes of
several satellite radii the flight times are rather several hours.
We therefore get an estimate of the total amount of dust contained
in a steady-state cloud around Ganymede: 1 to 100 tons.
Next, about $10^{-6}$ to $10^{-3}$ of the ejecta escape from
the satellite.
It gives the injection rate of the material into circumjovian
space of $\sim 10^{-4}$ to $1\kg\second^{-1}$, which may
be comparable with
the influx rate of IDPs to the Ganymede surface.

Remember that, in our data interpretation, we assumed no time
variations of the incoming impactor flux, so that the dust cloud
was assumed to be in a steady state. It would be interesting to
lift this assumption and to estimate possible time variations~---
\eg, temporary enhancements of number density caused by hits
of more massive meteorites, treated as individual events.
It should be noted, however, that no pronounced enhancements
of concentration {\em of micrometre-sized debris} are expected 
({\it Krivov and Jurewicz,} 1999)\nocite{krivov-jurewicz-1998}.

Another, and possibly more restrictive simplification
that we made is the spherical symmetry of the dust cloud,
which implies an isotropy of the impactor flux.
The mean speed of IDPs with respect to Jupiter at the distance of Ganymede
is about $25\km\second^{-1}$ (see Section~3).
Jupiter's motion about the Sun (with the velocity of $13\km\second^{-1}$)
and the orbital motion of the moon about Jupiter
($11\km\second^{-1}$) make the distribution of impactors at Ganymede
highly unisotropic.
Considering the geometry of the orbital motions of Jupiter, Ganymede, and
IDPs, it is easy to show that velocities of the grains with respect to
Ganymede vary from nearly zero to about $50\km\second^{-1}$,
depending on the position of Ganymede on its orbit and the directions
of the particle velocities.
As explained in Section 3, these extreme velocity values are attained when
Ganymede is at opposition to the Sun, at which moment we can expect a
very high asymmetry of the dust cloud above the leading and trailing
hemispheres of the moon.
The effect gets smaller at Ganymede's conjunction,
when the prograde particles hit the trailing hemisphere with $\ge 23$
and retrograde grains reach the leading side with $\le
27\km\second^{-1}$, but still many more impacts are expected on the
leading edge of Ganymede than on the trailing edge.
Therefore, we predict a marked and time-variable hemispheric
asymmetry.
Unfortunately, given the large statistical uncertainties due to the
small number of  detected particles, no obvious spatial and/or
temporal variation could be found in the present data.
For this reason, we do not undertake detailed modelling for the
angular distribution of impactors and ejecta similar to what has
been done in different problems 
(\eg, {\it Zook,} 1992; {\it Colwell}, 1993).\nocite{colwell-1993}
We hope to address these issues 
in future investigations of the full data set for all Galilean moons.

Much of the dust ejected from Ganymede is launched into
bound orbits and falls back to the moon.
A complex of these short-living, but continuously
replenished grains forms what we call an ejecta dust cloud
of Ganymede.
Obviously, all massive satellites which lack gaseous
atmospheres should own an ejecta dust cloud.    
Before Galileo, there were few attempts of direct in-situ
detections of large ejecta close to satellites~---
most notably, near the Moon
({\it Iglseder \etal,} 1996)\nocite{iglseder-et-al-1996};
these experiments did not bring up definite results, however.
Here we present the successful measurements of the satellite ejecta in
the vicinity of a source moon.  We also present
a dynamical model for an ejecta dust cloud of a massive satellite.

A tiny fraction of impact debris is ejected at speeds sufficient to
escape from Ganymede entirely. As shown above -- within the 
error bars -- the ejected mass is
comparable with the incoming flux of IDP 
imactors. The ejected material 
goes into orbit around Jupiter
and most of it will eventually be swept up by one of the Galilean satellites.
These escaping grains are probably responsible for some of the impact events
detected by the dust instrument in the inner Jovian system between the
Galilean satellites
({\it Gr\"un et al.,} 1997, 1998; {\it Krivov et al.} 2000).
Unfortunately, the ring of material formed by these grains escaping
from Ganymede is
far too tenuous to be detected optically. However, the fraction of
debris escaping a satellite is a steeply decreasing function of
satellite mass, so steep that despite their reduced cross sections,
small moons may be better sources of dust
than large satellites.
Indeed, many small moons have been proved to be,
or are supposed to be, sources of material for tenuous
dusty rings surrounding all giant planets
({\it Burns \etal,} 1984)\nocite{burns-et-al-1984}
and presumably Mars
({\it Soter,} 1971;\nocite{soter-1971}
{\it Krivov and Hamilton,} 1997).\nocite{krivov-hamilton-1997}
In the Jovian system, four innermost moons are sources
for the dust halo, the main ring, and the two-component gossamer ring
({\it Ockert-Bell \etal,} 1999;\nocite{ockert-bell-et-al-1998}
{\it Burns et al.,} 1999).
A variety of examples can be found in the Saturnian system,
which is especially important in view of the Cassini mission.
Enceladus supplies material to the huge E ring
({\it Hamilton and Burns,} 1994).\nocite{hamilton-burns-1994}
Irregularly-shaped icy Hyperion is believed to be a source of icy ejecta
which arrive at Titan, possibly affecting the complex chemistry
of its dense nitrogen atmosphere
({\it Banaszkiewicz and Krivov,} 1997;\nocite{banaszkiewicz-krivov-1997b}
{\it Krivov and Banaszkiewicz,} 2000).\nocite{krivov-banaszkiewicz-1999b}
Yet farther out from Saturn, the outermost retrograde moon Phoebe
emits dust which is thought to be deposited on the leading side of
Iapetus, producing its observed brightness
asymmetry ({\it Burns et al.,} 1996).\nocite{burns-et-al-1996}
All these cases exemplify the same mechanism of dust
production as the one discussed here for Ganymede.

Therefore, spacecraft measurements near the
satellites, \ie\ very close to the sources of dust, would be
of primary importance to gain more insight into the properties
of satellite surfaces and the dusty rings these moons maintain.
We believe the technique of processing and interpretation 
of spacecraft data which we proposed here and tested on Ganymede
can be directly applied to data from the dust experiment onboard the 
Cassini spacecraft which are to be obtained during its flybys 
of the Saturnian moons.

The Galileo dust measurements at Ganymede presented in this paper
can be considered as  a unique natural impact experiment. They 
complement laboratory experiments in an astrophysically relevant 
environment. Laboratory impact 
experiments have significant deficiencies in many respects, in the 
speeds of the projectiles and the mass and speed ranges in which 
ejecta particles can be observed. Furthermore, there is the 
always pending question of the astrophysical relevance of the 
materials used. Although far from being perfect impact
experiments, the Galileo results offer two extremely important
improvements over laboratory experiments: 1) the projectile 
and target materials and projectile speeds are astrophysically 
relevant, and 2) the masses and speeds of the ejecta particles can 
be determined in an important region of parameter space 
(micrometre sizes and $\rm km\,s^{-1}$ speeds).

\bigskip
{\bf Acknowledgements.}
We thank Andreas Heck and Valeri Dikarev for valuable discussions and 
the Galileo project at JPL for effective and successful
mission operations.
Thorough reviews by Herbert Zook, Markus Landgraf, Joshua Colwell,
and an anonymous referee are appreciated.
A.~K. greatly thanks his colleagues of the 
Heidelberg dust group for their warm hospitality and funding his 
stay at MPIK.
The final stage of the work was done during A.~K.'s Alexander von Humboldt
fellowship at MPAe.
This work has been supported by Deutsches
Zentrum f\"ur Luft- und Raumfahrt e.V. (DLR).


\section*{References}

\newcommand{\AAp}      {Astron. Astrophys.}
\newcommand{\AApT}     {Astron. Astrophys. Trans.}
\newcommand{\AdvSR}    {Advances in Space Research}
\newcommand{\AJ}       {Astron. J.}
\newcommand{\ApJ}      {Astrophys. J.}
\newcommand{\ApJL}     {Astrophysical Journal Letters}
\newcommand{\ApSS}     {Astrophysics and Space Science}
\newcommand{\ARAA}     {Annual Review in Astronomy and Astrophysics}
\newcommand{\BAAS}     {Bull. Am. Astron. Soc.}
\newcommand{\CelMech}  {Celestial Mechanics and Dynamical Astronomy}
\newcommand{\EMP}      {Earth, Moon, and Planets}
\newcommand{\EPS}      {Earth, Planets and Space}
\newcommand{\GRL}      {Geophysical Research Letters}
\newcommand{\JGR}      {J. Geophys. Res.}
\newcommand{\MNRAS}    {Monthly Notices Royal Astron. Soc.}
\newcommand{\PASJ}     {Publ. Astron. Soc. Japan}
\newcommand{\PASP}     {Publ. Astron. Soc. Pasific}
\newcommand{\PRL}      {Physical Review Letters}
\newcommand{\PSS}      {Planet. Space Sci.}
\newcommand{\SolPhys}  {Solar Physics}
\newcommand{\SolSysRes}{Solar System Research}
\newcommand{\SSR}      {Space Science Reviews}

{\small

Asada, N.
Fine fragments in high-velocity impact experiments.
{\em \JGR}, 90, 12,445--12,453, 1985.

Baguhl, M., Gr\"un, E., Linkert, G., Linkert, D., Siddique, N. 
and Zook, H., Identification of ''small'' dust impacts in the
Ulysses dust detector data, {\em Planet. Space Sci.}, 41, 1085-1098,
1993

Banaszkiewicz, M. and Krivov, A.~V.,
Hyperion as a dust source in the saturnian system.
{\em Icarus}, 129, 289--303, 1997.

Burchell, M.~J., {Brooke-Thomas}, W., {Leliwa-Kopystynski}, J.,
and Zarnecki, J.~C.,
Hypervelocity impact expreiments on solid $CO_2$ targets.
{\em Icarus}, 131, 210--222, 1998.

Burns, J.~A., Showalter, M.~R., and Morfill, G.~E.,
The ethereal rings of {Jupiter} and {Saturn}.
In Greenberg, R. and Brahic, A., Eds., {\em Planetary Rings},
  pp. 200--272. University of Arizona Press, Tucson, 1984.

Burns, J. A., Hamilton, D.~P., Mignard, F., and Soter, S.,
The contamination of {Iapetus} by {Phoebe} dust.
In Gustafson, B. A.~S. and Hanner, M.~S., Eds,
  {\em Physics, Chemistry, and Dynamics of Interplanetary Dust
  (ASP Conf. Series, vol. 104)},
  pp. 179--182. Kluwer, Dordrecht, 1996.

Burns, J. A., Showalter, M. R., Hamilton, D. P., Nicholson, P., D.,
de Pater, I., Ockert-Bell, M. E., Thomas, P., C., The formation
of Jupiter's faint rings, {\em Science}, 284, 1146-1150, 1999.

Colombo, G., Lautman, D.~A., and Shapiro, I.~I.
The {Earth's} dust belt: {Fact} or fiction? {2}. {Gravitational}
focusing and {Jacobi} capture.
{\em \JGR}, 71, 5,705--5,717, 1966.

Colwell, J.~E.,
A general formulation for the distribution of impacts and ejecta from
  small planetary satellites.
{\em Icarus}, 106, 536--548, 1993.

Colwell, J.~E. and Esposito, L.~W.
Origins of the rings of {Uranus} and {Neptune}. {II}~--- {Initial}
conditions and ring moon populations.
{\em \JGR}, 98, 7387--7401, 1993.

Colwell, J.~E. and Hor\'anyi, M.
Magnetospheric effects on micrometeoroid fluxes.
{\em \JGR}, 101, 2,169--2,175, 1996.

Colwell, J. E., Hor\'anyi, M., and Gr\"un, E.,
Capture of interplanetary and interstellar dust by the 
Jovian magnetosphere,
{\it Science}  280, 88--91, 1998.

Divine, N.
Five populations of interplanetary meteoroids.  
{\em \JGR}, 98, 17,029--17,048, 1993.

Frisch, W.,
Hypervelocity impact experiments with water ice targets.
In {McDonnell}, J. A.~M., Ed., {\em Hypervelocity Impacts in Space},
pp. 7--14. University of Kent, Canterbury, Great Britain, 1992.

Graps, A., Gr\"un, E., Svedhem, H., Kr\"uger, H., 
Hor\'anyi, M., Heck, A. and Lammers, S.,
Io as a source of the Jovian dust streams, {\em Nature}, 405, 48--50,
2000.


Gr\"un, E., Fechtig, H., Zook, H.~A., and Giese, R.~H.,
Collisional balance of the meteoritic complex.
{\em Icarus}, 62, 244--272, 1985.

Gr\"un, E., Fechtig, H., Hanner, M. S., Kissel, J., Lindblad, B.-A.,
Linkert, D., Linkert, G., Morfill, G. E., and Zook, H.,
The Galileo dust detector, {\em Space Sci. Rev.}, 60, 317-340, 1992.

Gr{\"u}n, E., Zook, H.~A., Baguhl, M., Balogh, A., Bame, S.~J., Fechtig, H.,
  Forsyth, R., Hanner, M.~S., Hor\'anyi, M., Kissel, J., Lindblad, B.-A.,
  Linkert, D., Linkert, G., Mann, I., McDonnell, J. A.~M., Morfill, G.~E.,
  Phillips, J.~L., Polanskey, C., Schwehm, G., Siddique, N., Staubach, P.,
  Svestka, J., and Taylor, A.
Discovery of jovian dust streams and interstellar grains by the
  {U}lysses spacecraft.
{\em Nature}, {\bf 362}, 428--430, 1993.

Gr\"un, E., Baguhl, M., Fechtig, H., Hamilton, D.P., Kissel, J.,
Linkert, D., Linkert, G.\ and Riemann, R.,
Reduction of Galileo and Ulysses dust data.
{\em Planet. Space Sci.} 43, 941-951, 1995.

Gr\"un, E., Kr\"uger, H., Dermott, S., Fechtig, H., Graps, A., 
Gustafson, B. A., Hamilton, D. P., Hanner, M. S., Heck, A., 
Hor\'anyi, M., Kissel, J., Lindblad, B.-A., Linkert, D., Linkert, 
G., Mann, I., McDonnell, J. A. M., Morfill, G. E., Polanskey, C., 
Schwehm, G., Srama, R. and Zook, H. A.
Dust measurements in the jovian magnetosphere. {\em Geophys. Res. Lett.}
24, 2171-2174, 1997.

Gr\"un, E., Kr\"uger, H., Graps, A., Hamilton, D. P., Heck, A.,
Linkert, G., Zook, H. A., Dermott, S., Fechtig, H., Gustafson, B. A.,
Hanner, M. S., Hor\'anyi, M., Kissel, J., Lindblad, B.-A., Linkert, D.,
Mann, I., McDonnell, J. A. M., Morfill, G. E., Polanskey, C., 
Schwehm, G., Srama, R.  Galileo Observes Electromagnetically 
Coupled Dust in the Jovian Magnetosphere. {\em J. Geophys. Res.}, 
103, 20,011-20,022, 1998.

Hamilton, D.~P. and Burns, J.~A.,
Origin of {Saturn's} {E} ring: {Self}-sustained, naturally.
{\em Science}, 264, 550--553, 1994.

Hartmann, W.~K.
Impact experiments. {1.} {Ejecta} velocity distributions and related
results from regolith targets.
{\em Icarus}, 63, 69--98, 1985.

Hor\'anyi, M., Burns, J.~A. and Hamilton, D.~P.
The dynamics of Saturn's E ring particles. 
{\em Icarus}, 97, 248-259, 1992.

Hor\'anyi, M., Morfill, G. and Gr\"un, E., 
Mechanism for the acceleration and ejection of dust grains from 
Jupiter's magnetosphere, {\em Nature}, 363, 144--146, 1993

Hor\'anyi, M. and Cravens, T. E., 
The structure and dynamics of Jupiter's ring,
{\it Nature}, 381, 293-295, 1996.

Iglseder, H., Uesugi, K., and Svedhem, H.,
Cosmic dust measurements in lunar orbit.
{\em \AdvSR}, 17(12), 177--182, 1996.

Kivelson, M.~G., Warnecke, J., Bennett, L., Joy, S., Khurana, K.~K.,
Linker, J.~A., Russell, C.~T., Walker, R.~J., and Polanskey, C.
{Ganymede}'s magnetosphere: {Magnetometer} overview.
{\em \JGR}, 103, 19,963--19,972, 1998.



Krivov, A.~V.,
On the dust belts of {Mars}.
{\em \AAp}, 291, 657--663, 1994.

Krivov, A.~V. and Banaszkiewicz, M.,
Unusual origin, evolution, and fates of icy ejecta from {Hyperion}.
{\em Icarus}, submitted, 2000.

Krivov, A.~V. and Hamilton, D.~P.,
Martian dust belts: {Waiting} for discovery.
{\em Icarus}, 128, 335--353, 1997.

Krivov, A.~V. and Jurewicz, A.,
The ethereal dust envelopes of the {Martian} moons.
{\em \PSS}, 47, 45--56, 1999.

Krivov, A.~V., Kr\"uger, H., Gr\"un, E., Thiessenhusen, K.-U. and
Hamilton, D. P.,
Evidence for a new dust ring of Jupiter between the Galilean satellites,
{\em Geophys. Res. Letters}, submitted, 2000.
 
Kr\"uger, H., Krivov, A. V., Hamilton, D. P., Gr\"un, E., 
Detection of an impact-generated dust cloud around Ganymede, 
{\em Nature}, 399, 558-560, 1999a.

Kr\"uger, H.,  Gr\"un, E., Graps, A., Lammers, S., Observations
of electromagneticlly coupled dust in the Jovian magnetosphere,
{\em Astrophys. and Space Sci.}, 264, 247 -- 256, 1999b.

Kr\"uger, H.,  Gr\"un, E., Hamilton, D. P., Baguhl, M., Dermott, 
S., Fechtig, H., Gustafson, B. A., Hanner, M. S., Heck, A.,
Hor\'anyi, M., Kissel, J., Lindblad, B.-A., Linkert, D., Linkert, G.,
Mann, I., McDonnell, J. A. M., Morfill, G. E., Polanskey, C., 
Riemann, R., Schwehm, G., Srama, R. and Zook, H. A., 
Three years of Galileo dust data: II. 1993 to 1995. 
{\em \PSS}, 47, 85--106, 1999c.

Kr\"uger, H.,  Gr\"un, E., Heck, A., Lammers, S., 
Analysis of the sensor characteristics of the Galileo dust detector 
with collimated Jovian dust stream particles, {\em \PSS},
47, 1015--1028, 1999d.

Kr\"uger, H.,  Gr\"un, E., Graps, A., Bindschadler, D., Dermott,
S., Fechtig, H., Gustafson, B. A., Hamilton, D. P., Hanner, M. S.,
Hor\'anyi, M., Kissel, J., Lindblad, B.-A., Linkert, D., Linkert, G.,
Mann, I., McDonnell, J. A. M., Morfill, G. E., Polanskey, C.,
Schwehm, G., Srama, R. and Zook, H. A.,
One year of Galileo dust data from the Jovian system: 1996.
{\em \PSS}, in prep., 2000.        


Landgraf, M., Baggeley, W. J., Gr\"un, E., Kr\"uger, H. and 
Linkert, G., Aspects of the mass distribution of interstellar dust 
grains in the solar system, {\em J. Geophys. Res.}, 105, No. A5, 10, 343,
2000.

Lange, M.~A. and Ahrens, T.~J.,
Impact experiments in low-temperature ice.
{\em Icarus}, 69, 506--518, 1987.


Morfill, G. E., Gr\"un, E., Johnson, T. V., 
Dust in Jupiter's magnetosphere: 
Origin of the ring, 
{\it Planet. Sp. Sci.}, 28, 1,101-1,110, 1980.

Nakamura, A. and Fujiwara, A.,
Velocity distribution of fragments formed in a simulated collisional
  disruption.
{\em Icarus}, 92, 132--146, 1991.

Ockert-Bell, M.~E., Burns, J.~A., Daubar, I.~J., Thomas, P.~C.,
Veverka, J.,  and Belton, M. A.~S.,
The structure of {Jupiter}'s ring system as revealed by the {Galileo}
  imaging experiment.
{\em Icarus}, 138, 188--213, 1999.

{O'Keefe}, J.~D. and Ahrens, T.
Impact and explosion crater ejecta, fragment size, and velocity.
{\em Icarus}, 62, 328--338, 1985.

Soter, S.,
The dust belts of {Mars}.
Report of Center for Radiophysics and Space Research No.~462, 1971.

St\"offler, D., Gault, D.~E., Wedekind, J., and Polkowski, G.
Experimental hypervelocity impact into quartz sand: {Distribution}
and shock metamorphism of ejecta.
{\em \JGR}, 80, 4042--4077, 1975.

Thiessenhusen, K.-U., 
Kr\"uger, H., Spahn, F., Gr\"un, E., 
Dust grains around Jupiter~--- the observations of the Galileo 
dust detector,
{\it Icarus}, 144, 89--98,  2000.

Zook, H. A., Gr\"un, E., Baguhl, M., Hamilton, D. P.,
Linkert, G., Liou, J.-C., Forsyth, R. and Phillips, J. L.,
Solar wind magnetic field bending of jovian dust trajectories,
{\em Science} 274, 1,501-1,503, 1996.
 
Zook, H. A.
Deriving velocity distribution of meteoroids from the measured meteoroid
impact directionality on the various LDEF surfaces.
In A. S. Levine, Ed., {\it LDEF -- 69 Months in Space
(First Post-Retrieval Symposium}, NASA CP-3134, Part 1, 569-579, 1992.

}



\newpage    

\newpage

\newpage


\newpage

\newpage

\newpage

\newpage

\newpage

\newpage

\newpage

\begin{table}[hb]
\caption{
Parameters for the dust particles detected within $\rm 10\,R_G$ altitude 
during Galileo's Ganymede flybys.
The first two columns list the flyby number and the time of the flyby
(year-day of year), column 3 gives the altitude at the closest approach
to Ganymede, column 4 gives the velocity of Galileo relative to Ganymede, 
column 5 lists the average particle velocity (only particles with 
a velocity error factor $\rm VEF < 6$ have been used, \cf\ 
{\it Gr\"un et al.} (1995), 
column 6 gives the number of class~2 and class~3 Ganymede
particles for which the complete data set has been transmitted to Earth, 
and column 7 gives the number of Ganymede particles after correction for 
incomplete data transmission and column 8 the slope of the power law fit 
to the radial variation of the impact rate (\cf\ Fig.~\ref{rate},
particles below $\rm 10\,R_G$ altitude have been included in the power law
fits for G1, G2 and G7, and below $\rm 6\,R_G$ for G8, respectively).}
\begin{tabular}{cccccccc}
\\
\hline
\hline 
Flyby&  Date    &Altitude & Spacecraft & Average    & Particles  & Corrected & Slope of \\
     &(year-day)&         & velocity   & particle   & with full  & number of & impact \\
     &          &         &            & velocity   & data set   & particles & rate   \\
     &          &   (km)  &
                    ($\km\second^{-1}$)& 
                                 ($\km\second^{-1}$)&            &            &     \\
  (1)&   (2)    &   (3)   &   (4)      &   (5)      &  (6)       &  (7)       & (8) \\
\hline
 G1  &96-179.270&\,\,\,\,\,844&  7.8     & $6.3 \pm 2.1$ &      15    &  30 & $-2.45\pm 0.16$\\
 G2  &96-250.791&\,\,\,\,\,262&  8.0     & $8.7 \pm 7.8$ &      10    &  48 & $-0.52\pm 0.10$\\
 G7  &97-095.299&  3,095  &    8.5     & $6.7 \pm 3.7$   &      11    &  11 & $-2.32\pm 0.02$\\
 G8  &97-127.665&  1,596  &    8.6     & $6.9 \pm 2.4$ &\,\,\,\,\,\,$9^{\ast}$&  49 & $\,\,\,\,0.37 \pm 0.22$\\
\hline
\hline
\end{tabular}
\mbox{}\\
\vspace{2mm}
*\small Only particles with impact velocity
v $\leq 10 \km\second^{-1}$ and below $\rm 6\,R_G$ altitude included.
\end{table}

\end{document}